\documentclass[preprintnumbers,amsmath,amssymb,floatfix,10pt,prd,onecolumn,
superscriptaddress,nofootinbib]{revtex4}
\usepackage{amsfonts}
\usepackage{mathrsfs}
\usepackage{latexsym}
\usepackage{epsfig}
\usepackage{epstopdf}
\usepackage{graphicx}
\usepackage{amssymb}
\usepackage{amsmath}
\usepackage{dcolumn}
\usepackage{bm}
\usepackage{color}
\usepackage{xcolor}
\usepackage{comment}
\usepackage[cp1251]{inputenc}
\begin{document}

\title{\bf Anisotropic Spheres Via Embedding Approach in $f(R,\phi,X)$ Gravity}

\author{Adnan Malik}
\email{adnan.malik@zjnu.edu.cn; adnanmalik_chheena@yahoo.com; adnan.malik@skt.umt.edu.pk}
\affiliation{School of Mathematical Sciences, Zhejiang Normal University, \\Jinhua, Zhejiang, China.}
\affiliation{Department of Mathematics, University of Management and Technology,\\ Sialkot Campus, Pakistan.}

\author{Yonghui Xia }
\email{xiadoc@163.com}
\affiliation{School of Mathematical Sciences, Zhejiang Normal University, \\Jinhua, Zhejiang, China.}

\author{Ayesha Almas}
\email{ayesha787@icloud.com}\affiliation{Department of Mathematics, University of Management and Technology,\\ Sialkot Campus, Pakistan.}

\author{M. Farasat Shamir}
\email{farasat.shamir@nu.edu.pk}
\affiliation{National University of Computer and Emerging Sciences,\\ Lahore Campus, Pakistan.}



\begin{abstract}
\begin{center}
\textbf{Abstract}\\
\end{center}

\bigskip
In this manuscript, we investigate the behavior of stellar structure through embedding approach in $f(R, \phi, X)$ modified theory of gravity, where $R$ denotes the Ricci scalar, $\phi$ represents the scalar potential and $X$ indicates the kinetic potential. For this purpose, we consider the spherically symmetric space-time with anisotropic fluid. We further choose three different stars i.e. LMC X-4, Cen X-3, and EXO 1785-248 to demonstrate the behavior of stellar structures. We further compare the Schwarzschild space-time as exterior geometry with spherically symmetric space-time to calculate the values of unknown parameters. In this regard, we investigate the graphical features of stellar spheres such es energy density, pressure components, anisotropic component, equation of state parameters, stability analysis and energy conditions. Furthermore, we investigate some extra conditions such as mass function, compactness factor and surface redshift respectively. Conclusively, all the compact stars under observations are realistic, stable, and are free from any physical or geometrical singularities. We find that the embedding class one solution for anisotropic compact stars is viable and stable. \\\\
{\bf Keywords:} Anisotropic Spheres, Compact Stars, Embedding Approach, $f(R,\phi,X)$ Gravity.
\end{abstract}
\maketitle

\date{\today}

\section{Introduction}
Recent explanations based on astrophysical facts have revealed an amazing picture of the expansion of the universe \cite{1a,2a,3a}. Cosmologist have brought out new ideas to introduce the critical innovations for this accelerated expanding universe. There are two types of serious issues in cosmological models known as dark energy and dark matter. The dark matter is considered to occupy a major portion of this accelerating universe and it is the main cause of acceleration of universe. It is expected that the modification in the theory of general relativity may explain the accelerating expansion of universe. However, this approach has some limitations because dark energy has never been directly detected or observed. The concept of dark energy, which is never detected experimentally, is also a one way for solving this problem. According to Sloan Digital Sky Surveys (SDSS) \cite{4a}, the BICEP2 experiment \cite{5a,6a,7a}, Planck satellite \cite{8a,9a,10a} and  the Wilkinson Microwave anisotropy probe (WMAP) \cite{11a,12a}, it happen that 27\% of universe is composed of the dark matter, 68\% is of dark energy and the rest is ordinary matter. The modification in Einstein theory of general relativity seems to be a good approach to justify the idea of dark energy. Through this approach, many models are introduced for explaining the universe expansion and observing some interesting aspects of nature. It has led to a search for modified or extended gravitational theories capable of addressing such challenges. General relativity as a physical theory has been a great success in the previous century but there are still some issues which could not be addressed properly like dark energy, dark matter, initial singularity, late-time cosmic acceleration and flatness problems. Some alternative models of gravity are proposed which are believed to be a real cause of this accelerating expansion of the universe. Several modified theories have suggested as alternatives to general relativity in recent years. Some of these gravitational theories include $f(R)$ \cite{4.01,4.02,4.1,4.2,4.3}, $f(R, G)$ \cite{4.4}, $f(G)$ \cite{4.5,4.6,4.7,4.8}, $f(R, T)$ \cite{4.09,4.9,4.9a,4.9b}, $f(R ,\phi)$ \cite{4.10,4.11,4.11a,4.11b} and $f(R ,\phi, X)$ \cite{4.12,4.13} modified theories of gravity. These modified theories of gravity explain the phenomena of expansion of universe. These theories explain the weak field regimes but some modifications are still required to address the strong field for universe expansion.\\

 The research for precise static spherically solutions for relativistic structures is a difficult problem to solve because of the close association of non-linear elements in the modified Einstein-Maxwell equations. To solve this problem, we may use the embedding class I technique with the Eishenhart condition to identify new physically acceptable solutions for the sphere's compact geometry. New anisotropic results may be generated from a perfectly distributed fluid in a direct, ordered, and simple manner. Regarding the creation of novel spherical solutions, the embedding class I technique exhibits a vast array of noteworthy components. Nazar and Abbas \cite{a14} investigated a class of an exact analytical solutions of shear-free and spherically symmetric gravitational collapse of Karmarkar star in the minimally coupled $f(R)$ theory of gravity by incorporating the properties of anisotropic radiating matter. Malik and his collaborators \cite{a16} used an embedded class-I technique to show the evolution of anisotropic stellar formations against the framework of $f(R)$ modification of gravity. Maurya et al., \cite{a17} presented a hierarchical solution-generating technique employing the minimum gravitational decoupling method and the generalized concept of Complexity as applied to Class I spacetime for bounded compact objects in classical general relativity. Sharif and Naseer \cite{a18} studied several specific anisotropic stellar models in $f(R, T, Q)$ modified gravity by employing spherically symmetric configuration to create solutions of modified field equations corresponding to distinct matter Lagrangian options using the embedding class-one method. Gudekli and colleagues \cite{a19} investigated the spherically symmetric solutions of embedding class-one in the $f(T, \tau)$ theory of gravity, providing an extended compact star model. Sarkar et al., \cite{a20} suggested an entirely new model for spherically symmetric anisotropic astrophysical objects with class I solutions in the framework of $f(R, T)$ gravity. Errehymy et al., \cite{a21} examined the presence of compact objects characterizing anisotropic matter distributions in the context of $f(R, T)$ gravity and utilize the embedding class-I approach to generate a comprehensive space-time interpretation on the inside of the stellar structure.\\

Ditta and Xia \cite{a22} investigated stellar formations using the Karmarkar condition and Tolman-Kuchowiz metric components with an anisotropic fluid allocation within the context of Rastall teleparallel gravity. Sharif and Naseer \cite{a23} examined the charged stellar models linked with an anisotropic source of fluid allocation in $f(R, T, Q)$ gravity by analyzing a self-gravitational spherical configuration in the involvement of an electromagnetic field and generating solutions to the field equations by employing the Karmarkar condition and the equation of state (MIT bag model). Sharif and Hassan \cite{a25} addressed the methodology of a complexity factor for a dynamical anisotropic stellar structure in the context of $f(G, T)$ modified gravity and analyzed the formation scalars by orthogonal division of the Riemann tensor to calculate a complexity factor that includes all the basic principles of the structure. Zubair et al., \cite{a26} investigated the generalized symmetric, static compact objects under anisotropic fluid in the background of Karmarkar embedding class-1 condition by considering the gravitational Lagrangian as a linear function of the Ricci scalar and the trace of the stress–energy tensor. Using Pant's inner solution, Pant et al., \cite{a27} investigated a novel embedding of an anisotropic charged form of a solution to modified field equations in the 4-dimentional configuration using the Karmarkar conditions and the gravitational decoupling through the minimum geometric decoupling approach. Usman and Shamir \cite{a28} examined the effects of gravitational breakdown by looking at heat flux anisotropic sources of fluid in the context of $f(R)$ modified gravity and employing the non-static spherically symmetric spacetime to describe the essence of the internal spacetime and compare it with the Vaidya external configuration.\\

In this work, we intend to consider a generalized modified gravity theory, $f(R, \phi, X)$ gravity, where $R$ is the Ricci scalar, $\phi$ a scalar field and $X$ a kinetic term. This theory contains a wide range of known dark energy and modified gravity models, for instance $f(R)$ gravity models or Galileons. In particular, several cosmological solutions are studied within the framework of these theories, specifically solutions that can provide cosmic acceleration at late times, and even the exact $\Lambda$CDM evolution. Reconstruction techniques are implemented in order to obtain the $f(R, \phi, X)$ of the action given a particular Hubble parameter. This provides a way to efficiently check the viability of any gravitational action by just considering a particular cosmological evolution and then analyzing the gravitational action. An action of the form $f(R, \phi, X)$ is in fact quite natural, as it removes any assumptions on the underlying theory of gravity with the exception of being second order. We can think of the field $\phi$ as the effective field controlling the strength of the gravitational force. Bahamonde et al. \cite{ad1} explored a generalized $f(R, \phi, X)$ modified gravity theory, which contains a wide range of dark energy and modified gravity models. They also considered specific models and applications to the late-time cosmic acceleration. Bahamonde along with his collaborators \cite{ad2} investigated new exact spherically symmetric solutions in $f(R, \phi, X)$ theory of gravity by Noether’s symmetry approach, and some of these solutions can represent new black holes solutions in this extended theory of gravity. Malik, et al. \cite{ad3} investigated the behavior of anisotropic compact stars in generalized modified gravity, namely $f(R, \phi, X)$ by considering the spherically symmetric spacetime to analyze the feasible exposure of compact stars. Malik along with his collaborators \cite{ad4} investigated some cylindrically symmetric solutions in a very well known modified theory named as $f(R, \phi, X)$ by taking the cylindrically symmetric space-time to discuss the cylindrical solutions in some realistic regions. Shamir, et al. \cite{ad5} considered the spherically symmetric static spacetime with an anisotropic fluid source to discuss the wormhole solutions by using two well-known distributions like Gaussian and Lorentzian non-commutative geometry in $f(R, \phi, X)$ theory of gravity. The same authors \cite{ad5} considered a particular equation of state parameter to study the behavior of traceless fluid and examined the physical behavior of wormhole solutions in the background of $f(R, \phi, X)$ theory of gravity. Recently, Malik et al., \cite{ad7} investigated the concept of cracking and overturning to analyze the impact of local density perturbations on the stability of self-gravitating compact objects in the framework of $f(R, \phi, X)$ theory of gravity. Recently, Malik along with his collaborators \cite{ad8} provided a new model of anisotropic strange star corresponding to the exterior Schwarzschild metric and the Einstein field equations have been solved by utilizing the Krori-Barua ansatz in $f(R, \phi, X)$ theory of gravity. Malik et al., \cite{ad9} investigated and analyzed the behavior of charged compact stars in the modified $f(R, \phi, X)$  theory of gravity by assuming the Krori– Barua space-time. \\

To the best of our knowledge, no attempt has been made so far to discuss the spherically symmetric solutions of embedding class I technique in $f(R,\phi, X)$ theory of gravity. In this paper, we are inspired to investigate the nature of stellar structure in the $f(R, \phi, X)$ theory of gravity utilizing the Karmarkar condition. The arrangement of this paper is as follows: Section II deals with the field equations of $f(R,\phi, X)$ gravity in the presence of embedding class I. In Section III, we investigate the matching conditions for finding the unknown parameters. All the graphical representation of the stellar configuration has been discussed in Section IV. The last section deals with the concluding remarks.

\section{Basic Formulism of $f(R,\phi,X)$ gravity}
The action for the modified $f(R,\phi,X)$ theory of gravity \cite{ad10,ad11} is given as
\begin{equation}\label{1}
S=\int\sqrt{-g}\bigg(\textit{L}_m+\frac{1}{2\kappa^2}f(R,\phi,X)\bigg)d^4x,
\end{equation}
where, $g$ represents the determinant of $g_{\eta\xi}$, $\textit{L}_m$ is lagrangian matter, and $X$ is kinetic term, which is defined as
\begin{equation}\label{2}
X=\frac{-\varepsilon}{2}\partial^\nu\phi\partial_\nu\phi.
\end{equation}
Here, $\varepsilon$ is a parameter. By applying variation to Eq. (\ref{1}) with respect to $g_{\eta\xi}$, we get the field equation as
\begin{equation}\label{3}
f_RG_{\eta\xi}-\frac{1}{2}(f-Rf_R)g_{\eta\xi}-\nabla_\eta\nabla_\xi f_R+g_{\eta\xi}\nabla_\nu\nabla^\nu f_R-\frac{1}{2}f_X(\nabla_\eta \phi)(\nabla_\xi \phi)=\kappa^2 T_{\eta\xi}.
\end{equation}
For our convenience, we take $f\equiv f(R,\phi,X)$ and $f_R$ is a partial derivative of $f$ w.r.t $R$. Whereas, $\nabla_\nu$ is a covariant derivative and  $T_{\eta\xi}$ is the energy-momentum tensor (EMT) \cite{ad12} yields the following expression,

\begin{equation}\label{5}
T_{\eta\xi} = (\rho+p_{t})\upsilon_{\eta}\upsilon_{\xi}-p_{t}g_{\eta\xi}+(p_{r}-p_{t})\vartheta_{\eta}\vartheta_{\xi},
\end{equation}
where, $\upsilon_{\eta}$ and $\vartheta_{\xi}$ are four vector velocity components, i.e., $\upsilon_{\eta}= e^{\frac{\lambda}{2}}\delta^0_\eta$ and $\vartheta_{\xi}=e^{\frac{\zeta}{2}}\delta^1_\xi$. Whereas, $\rho$, $p_r$, and $p_t$ represent the energy density, radial pressure and tangential pressure respectively. This anisotropy feature influences the physical properties, such as gravitational redshift, energy density, and total mass \cite{ad13}. Moreover, theoretical studies indicate that the pressure within compact stars with extreme internal density and strong gravity may be anisotropic \cite{ad13a}. We consider the spherically symmetric spacetime as
\begin{equation}\label{6}
ds^{2} = e^{\lambda(r)}dt^{2}-e^{\zeta(r)}dr^{2}-r^{2}(d\theta^{2}+\sin^2\theta d\phi^{2}),
\end{equation}
where, $\lambda$ and $\zeta$ are the function of radial coordinate $r$. For further analysis, we choose the following model of $f(R,\phi,X)$ gravity \cite{ad7} as
\begin{equation}\label{7}
f(R,\phi,X)=R +\gamma R^2+X-V(\phi).
\end{equation}
The model defined in Eq. (\ref{7}) often referred to particular well-known Starobinsky $R+\gamma R^2$ model, has been proposed as an alternative to explain various astrophysical phenomena including compact object. We choose this model because it shows exponential growth for early-time cosmic expansion. By adding term like $\gamma R^2$ introduce modification to the standard general relativity that can potentially account for gravitational effect not explained by general relativity \cite{ad14}. Moreover, this term interpolate a curvature dependence that affect gravitational dynamics also it allows the possibility of a non-linear relation between the spacetime curvature and the field due to gravity. It is likely to mention here that if the value of parameter $\gamma$ chosen to be non-positive then beyond a maximum mass limit is reached. However, this give rise to a problem, particularly, the Ricci scalar $R$ exhibits damped oscillation. Conversely, as we move towards infinity with non-negative values of $\gamma$, the R gradually decreases to zero, leading to a maximum mass for the star that is lower then $2M/M_{\odot}$. Furthermore, the term $V(\phi)=w_0\phi^n$ including $w_0$ and $n$ are arbitrary non-zero constant, which illustrates a potential energy function, which can be related to concept such as dark energy or inflation. As far as the viability of the model in terms of solar system tests is concerned, it is a debateable issue and require some extensive analysis. However, it has been shown that the $f(R)$ modified theories may pass the solar system observational constraints even if the scalar field is added \cite{ad1}. This shows that our consider $f(R,\phi,X)$ gravity model may pass solar system tests. Now using Eq. (\ref{6}) along with Eq. (\ref{5}) and (\ref{7}) in Eq. (\ref{3}), we get the following equations:
\begin{equation}\label{8}
\begin{split}
\rho=&\frac{1}{8r^4}e^{-2\zeta(r)}(8(-1+e^{\zeta(r)})(e^{\zeta(r)}(r^2-2\gamma)+10\gamma)-2e^{\zeta(r)}r^{2+2\alpha}\alpha^2+4e^{2\zeta(r)}r^4(r^\alpha)^nw_0+r(r^3\gamma\lambda^{'}(r)^4+8e^{\zeta(r)}(r^2-8\gamma)\\ \\ &\zeta^{'}(r)-2r^2\gamma\lambda^{'}(r)^3(-4+r\zeta^{'}(r))+r\gamma\lambda^{'}(r)^2(16+r(8\zeta^{'}(r)-3r\zeta^{'}(r)^2+4r(\lambda^{''}(r)+2\zeta^{''}(r))))+4r\gamma\lambda^{'}(r)(-8r\zeta^{'}(r)^2\\ \\ &+r^2\zeta^{'}(r)^3+\zeta^{'}(r)(4+5r^2(\lambda^{''}(r)-\zeta^{''}(r)))+2r(-2\lambda^{''}(r)+6\zeta^{''}(r)-2r\lambda^{'''}(r)+r\zeta^{'''}(r)))-4r\gamma(-4r\zeta^{'}(r)^3+\zeta^{'}(r)^2\\ \\ &(12+5r^2\lambda^{''}(r))-8\zeta^{''}(r)-4r\zeta^{'}(r)(6\lambda^{''}(r)-5\zeta^{''}(r)+2r\lambda^{'''}(r))+r(3r\lambda^{''}(r)^2-8r\lambda^{''}(r)\zeta^{''}(r)+16\lambda^{'''}(r)-8\zeta^{'''}(r)\\ \\&+4r\lambda^{''''}(r))))),
\end{split}
\end{equation}
\begin{equation}\label{9}
\begin{split}
p_r=&\frac{1}{8r^4}e^{-2\zeta(r)}(-8(-1+e^{\zeta(r)})(e^{\zeta(r)}(r^2-2\gamma)-14\gamma)-2e^{\zeta(r)}r^{2+2\alpha}\alpha^2-r(4e^{2\zeta(r)}r^3(r^{\alpha})^nw_0+r^3\gamma\lambda^{'}(r)^4+2r^3\gamma\lambda^{'}(r)^3\\ \\ &\zeta^{'}(r)+r^2\gamma\lambda^{'}(r)^2(24\zeta^{'}(r)+5r\zeta^{'}(r)^2+4r(-\lambda^{''}(r)+\zeta^{''}(r)))-4\gamma(8r^2\zeta^{'}(r)^3-6r\zeta^{'}(r)^2(-2+r^2\lambda^{''}(r))+r(16\lambda^{''}(r)\\ \\ &-r^2\lambda^{''}(r)^2-16\zeta^{''}(r)+8r\lambda^{'''}(r))+4\zeta^{'}(r)(-4+4e^{\zeta(r)}-2r^2\zeta^{''}(r)+r^3\lambda^{'''}(r)))+8\lambda^{'}(r)(-e^{\zeta(r)}r^2+\gamma(8+r(r\zeta^{'2}\\ \\ &-r^2\zeta^{'}(r)^3+\zeta^{'}(r)(12+r^2(-\lambda^{''}(r)+\zeta^{''}(r)))-r(6\lambda^{''}(r)-4\zeta^{''}(r)+r\lambda^{'''}(r))))))),
\end{split}
\end{equation}
\begin{equation}\label{10}
\begin{split}
p_t=&\frac{1}{8r^4}e^{-2\zeta(r)}(-16(-1+e^{\zeta(r)})(7+e^{\zeta(r)})\gamma+2e^{\zeta(r)}r^{2+2\alpha}\alpha^2-4e^{2\zeta(r)}r^4(r^{\alpha})^nw_0+r(r^3\gamma\lambda^{'}(r)^4+2r^2\gamma\lambda^{'}(r)^3(2-3r\zeta^{'})\\ \\ &+4e^{\zeta(r)}(-((r^2-4\gamma)\zeta^{'}(r))+r^3\lambda^{''}(r))+r\lambda^{'}(r)^2(2e^{\zeta(r)}r^2+\gamma(-16+r(-32\zeta^{'}(r)+9r\zeta^{'}(r)^2+12r(\lambda^{''}(r)-\zeta^{''}))))\\ \\ &+2\lambda^{'}(r)(e^{\zeta(r)}(2r^2+24\gamma-r^3\zeta^{'}(r))-2\gamma(4+r(-11r\zeta^{'}(r)^2+r^2\zeta^{'}(r)^3+\zeta^{'}(r)(-4+5r^2(2\lambda^{''}(r)-\zeta^{''}(r)))+2r(-\\ \\ &5\lambda^{''}(r)+7\zeta^{''}(r)-3r\lambda^{'''}(r)+r\zeta^{'''}(r)))))+4\gamma(-4r^2\zeta^{'}(r)^3+5r^3\zeta^{'}(r)^2\lambda^{''}(r)+4\zeta^{'}(r)(3+r^2(-5\lambda^{''}(r)+5\zeta^{''}(r)-\\ \\ &2r\lambda^{'''}(r)))+r(5r^2\lambda^{''}(r)^2-8\lambda^{''}(r)(1+r^2\zeta^{''}(r))+4r(3\lambda^{'''}(r)-2\zeta^{'''}(r)+r\lambda^{''''}(r)))))).
\end{split}
\end{equation}
The above Eqs. (\ref{8})-(\ref{10}) are very complex and non-linear differential equation. Now, we explore a major tool of the present study, which is the Karmarkar condition \cite{ad14}, suggested by Karmarkar. By using this, we apply an embedding technique to get all possible embedding class one spherically symmetric solutions. Eishenhart \cite{ad15} proposes the following necessary and significant conditions for second-order $\chi_{\eta\omega}$ tensor and Riemann tensor $\textrm{\textit{\textsf{R}}}_{\eta\xi\omega m}$:
\begin{equation}\nonumber
\textrm{\textit{\textsf{R}}}_{\eta\xi\omega m}=\Sigma(\chi_{\eta\omega}\chi_{\xi m} - \chi_{\eta m}\chi_{\xi\omega}),~~~~~~~~\chi_{\eta\xi;\omega} -  \chi_{\eta\omega;\xi} = 0.
\end{equation}
Here, $\Sigma= \pm1$. All the Riemann tensor for embedded class one given in the above equation are as follows:
\begin{equation}\nonumber
 \textrm{\textit{\textsf{R}}}_{1414} = \frac{e^{\lambda(r)} (2\lambda^{''}(r) + \lambda^{'}(r)^{2} - \lambda^{'}(r)\zeta^{'}(r))}{4},~~~ \textrm{\textit{\textsf{R}}}_{2323} = \frac{r^{2}\sin^{2}\theta (e^{\zeta(r)} - 1)}{e^{\zeta(r)}}, ~~~\textrm{\textit{\textsf{R}}}_{1334} = \textrm{\textit{\textsf{R}}}_{1224} \sin^{2}\theta,
\end{equation}
\begin{equation}\nonumber
\textrm{\textit{\textsf{R}}}_{1212} = \frac{r\zeta^{'}(r)}{2},~~~~~ \textrm{\textit{\textsf{R}}}_{3434} = \frac{r\sin^{2}\theta \zeta^{'}(r) e^{\lambda(r) -\zeta(r)}}{2},~~~~~\textrm{\textit{\textsf{R}}}_{1224} = 0.
\end{equation}
Now, Karmarkar's condition is described as
\begin{equation}\label{11}
   \textrm{\textit{\textsf{R}}}_{1414} \textrm{\textit{\textsf{R}}}_{2323} = \textrm{\textit{\textsf{R}}}_{1224} \textrm{\textit{\textsf{R}}}_{1334} + \textrm{\textit{\textsf{R}}}_{1212} \textrm{\textit{\textsf{R}}}_{3434} , ~~~\textrm{\textit{\textsf{R}}}_{2323}\neq 0.
\end{equation}
Pandey and Sharma \cite{abc1} acquired the following symmetric spacetime
\begin{equation}\label{12}
ds^{2} = e^{\lambda}dt^{2}-dr^{2}-r^{2}(d\theta^{2}+\sin^2\theta d\phi^{2}).
\end{equation}
which fulfills Karmarkar condition, still, is not embedded class one because $\textrm{\textit{\textsf{R}}}_{2323}= 0$. They argued that in symmetric spacetime, Karmarkar condition does not suffice to constitute a class one model. Therefore, the exact solution of field equations in the case of metric (\ref{6}) can be restricted as class one model if it sufficient (\ref{11}) along with $\textrm{\textit{\textsf{R}}}_{2323}\neq 0$. Now, by using Eq. (\ref{11}), we get a following differential equation as
\begin{equation}\label{13}
\frac{\lambda^{'} \zeta^{'}}{1 -e^{\zeta}}-(\lambda^{'} \zeta^{'} +\lambda^{'^2} -2(\lambda^{''} +\lambda^{'^2})) = 0,
\end{equation}
where, ${e^{\zeta} \neq 1}$. By integrating Eq. (\ref{11}), we get link between two metric potential components of the line element as
\begin{equation}\label{14}
e^{\zeta} =  e^{\lambda} \lambda^{'^2}+1 +D,
\end{equation}
where $D$ is an integration constant. Now, $g_{tt}$ can be chosen as
\begin{equation}\label{15}
e^{\lambda} = \varrho(1 +hr^{2})^{\tau}.
\end{equation}
where, $\tau$ is a positive integer, whereas $\varrho$ and $h$ are parameters. It is mentioned that at $r \rightarrow 0$, $e^{\lambda(r)} = \varrho$, which shows that the metric potential, we have selected is finite and regular at the core. Similarly, Lake \cite{abc2} demonstrated that a metric potential $e^{\lambda}$ must be non-negative and increasing monotonically for any physically viable stellar composition throughout the configuration. Hence, the metric potential we assume by (\ref{15}) is obeying all the mandatory requirement. Now, by substituting Eq. (\ref{15}) in (\ref{14}), we get the $g_{rr}$ component given as
\begin{equation}\label{16}
e^{\zeta(r)} = 1 +ohr^{2} (1 +hr^{2})^{-2 +\tau},
\end{equation}
where ${o = 4\tau^{2}h\varrho D}$. Now, by plugging Eq. (\ref{15}) and Eq. (\ref{16}) into Eqs. (\ref{8})-(\ref{10}), we obtain the following set of equations for the stellar configuration as
\begin{equation}\label{17}
\begin{split}
\rho=&\frac{1}{4r^2t_2(r)^5}\bigg(4o^5h^5r^8t_1(r)^{5\tau}(r^2-2\gamma)+8\tau h^2r^2t_1(r)^6t_3(r)\gamma+r^{2\alpha}t_1(r)^{10}\alpha^2+4o^3h^3r^4t_1(r)^{1+3\tau}(6(2+\tau hr^2)(r+hr^3)^2\\ \\&+2t_4(r)\gamma-r^{2+2\alpha}t_1(r)^3\alpha^2)+4ohr^2t_1(r)^{4+\tau}(3-r^{2\alpha}\alpha^2+2h^2r^2t_5(r)+2ht_6(r)+2h^3r^4t_7(r)+h^4r^6t_8(r))o^4h^4r^6\\ \\&t_1(r)^{1+4\tau}(-56\gamma+r^2(24+8h((1+\tau)r^2+(9-8\tau)\gamma)-r^{2\alpha}t_1(r)\alpha^2))+2o^2h^2r^2t_1(r)^{2+2\tau}(-276\gamma+r^2t_9(r))+2r^2\\ \\&(r^{\alpha})^{n}(t_1(r)^2+ohr^2t_1(r)^{\tau})^5w_{0}\bigg),
\end{split}
\end{equation}
\begin{equation}\label{18}
\begin{split}
p_r =&\frac{1}{4r^4t_2(r)^5}\bigg(8t_1(r)^4t_{10}(r)t_{11}(r)\gamma t_2(r)-128t_1(r)^6t_{10}(r)^3(2+(2+\tau)hr^2)\gamma-8t_1(r)(-21+hr^2t_{12})\gamma t_2(r)^3-32t_1^{2}\\ \\&(30+hr^2t_{13}(r))\gamma t_2(r)^2+t_1(r)(112\gamma+r^2t_{14}(r))t_2(r)^4-2r^4(r^{\alpha})^{n}w_{0}-4r^2+8\gamma\bigg),
\end{split}
\end{equation}
\begin{equation}\label{19}
\begin{split}
p_t=&\frac{1}{4r^2t_2(r)^5}\bigg(-8o^5h^5r^8t_1(r)^{5\tau}\gamma+8\tau^4h^4r^6t_1(r)^2(t_1(r)^4+16o^2h^2r^4t_1(r)^{2\tau}+23ohr^2t_1(r)^{2+\tau}\gamma)+32\tau^3h^3r^4t_1(r)^2\\ \\&(-((-2+hr^2)t_1(r)^4)+120^2h^2h^2r^4t_1(r)^{2\tau}+9ohr^2(4+hr^2)t_1(r)^{2+\tau})\gamma+4\tau^2h^2r^2t_1(r)^2t_{15}(r)+8\tau hr^2t_1(r)t_{16}(r)\\ \\&+r^{2\alpha}t_1(r)^{10}\alpha^2+4ohr^2t_1(r)^{5+\tau}(-1+hr^2(-1+hr^2+h^2r^4)+32h(5+hr^2(-17+2hr^2))\gamma+r^{2\alpha}t_1(r)^3\alpha^2)+2o^2\\ \\&h^2r^2t_1(r)^{3+2\tau}(6t_1(r)(r+hr^3)^2+4(37+hr^2(261+hr^2(89+79hr^2)))\gamma+3r^{2+2\alpha}t_1(r)^3\alpha^2)+o^4h^4r^6t_1(r)^{1+4\tau}(72\gamma\\ \\&+r^2(4+4h(r^2+26\gamma)+r^{2\alpha}t_1(r)\alpha^2))+4o^3h^3r^4t_1(r)^{1+3\tau}t_{17}(r)-2r^2(r^{\alpha})^{n}(t_1(r)^2+ohr^2t_1(r)^{\tau})^5w_0\bigg).
\end{split}
\end{equation}
where $t_{\textit{i}}(r)$, \{$\textit{i}=1,2,...,17$\}, are given in the Appendix. Whereas the expressions of the parameters $h,$ $o$, $\varrho$ will be determine from the matching condition.

\begin{table}[ht]
\centering
\caption{Values of $M$, $a$, $c_1$, $\rho_c$ and $p_c$ for LMC X-4 Stars with $R=10~ km$.}
\begin{tabular}{|c|c|c|c|c|}
  \hline
  $\tau$ & $o$ & $h$ & $\varrho$ & $\gamma$ \\
  \hline
  3 &~~4.7820~~&~~0.001273~~&~~0.4329~~&~~0.05584110  \\
  \hline
  5 &~~7.6640~~&~~0.000729~~&~~0.4363~~&~~0.03277414\\
  \hline
  10 &~~14.900~~&~~0.000352~~&~~0.4387~~&~~0.01556560\\
  \hline
  20 &~~29.392~~&~~0.000173~~&~~0.4399~~&~~0.00702774\\
  \hline
  50 &~~72.882~~&~~0.000068~~&~~0.4406~~&~~0.00702774\\
  \hline
  100 &~~145.373~~&~~0.000034~~&~~0.4408~~&~~0.00023905\\
  \hline
  500 &~~725.285~~&~~$0.000007$~~&~~0.4410~~&~~0.00011337\\
  \hline
\end{tabular}
\label{tab1}
\end{table}

\begin{table}[ht]
\centering
\caption{Numerical values of $o$, $\varrho$, $h$, and $\gamma$ for Cen X-3 star model 2.}
\begin{tabular}{|c|c|c|c|c|}
  \hline
$\tau$ & $o$ & $h$ & $\varrho$ & $\gamma$ \\
 \hline
3 &~~4.5683~~&~~0.001421~~&~~0.3765~~&~~0.05584110  \\
5 &~~7.2751~~&~~0.000805~~&~~0.3807~~&~~0.03277414\\
10 &~~14.082~~&~~0.000386~~&~~0.3838~~&~~0.01556560\\
20 &~~28.267~~&~~0.000186~~&~~0.3852~~&~~0.00702774\\
50 &~~68.656~~&~~0.000074~~&~~0.3861~~&~~0.00702774\\
100 &~~138.251~~&~~0.000036~~&~~0.3864~~&~~0.00023905\\
500 &~~682.743~~&~~$7.447003\times10^{-6}$~~&~~0.3866~~&~~0.00011337\\
 \hline
\end{tabular}
\label{tab1}
\end{table}

\begin{table}[ht]
\centering
\caption{Numerical values of $o$, $\varrho$, $h$, and $\gamma$ for EXO 1785-248 star model 3.}
\begin{tabular}{|c|c|c|c|c|}
  \hline
$\tau$ & $o$ & $h$ & $\varrho$ & $\gamma$ \\
\hline
3 &~~4.5698~~&~~0.001862~~&~~0.3768~~&~~0.07644824 \\
5 &~~7.2777~~&~~0.001055~~&~~0.3811~~&~~0.04422569\\
10 &~~14.361~~&~~0.000496~~&~~0.3842 ~~&~~0.02079960\\
20 &~~27.733~~&~~0.000248~~&~~0.3856~~&~~0.0093616181\\
50 &~~68.685~~&~~0.000098~~&~~0.3864~~&~~0.00259253\\
100 &~~136.944~~&~~0.000048~~&~~0.3867~~&~~0.00035201\\
500 &~~683.029~~&~~$9.759977\times10^{-6}$~~&~~0.3869~~&~~0.00014346\\
\hline
\end{tabular}
\label{tab1}
\end{table}

\section{Matching Conditions}
In order to find the values of parameters {$o$, $\varrho$, $h$}, which depict the model of relativistic anisotropic sphere,  we compare the spherically symmetric space-time with the Schwarzschild metric as exterior space-time and develop some matching conditions for finding the solutions. The external Schwarzschild spacetime is defined as
\begin{equation}\label{20}
ds^{2} = \left(1 -\frac{2M}{r}\right )dt^{2} -\left(1 -\frac{2M}{r}\right )^{-1} dr^{2} -r^{2}d\theta^{2} -r^{2}\sin^{2}\theta d\phi^{2},
\end{equation}
where `$M$' represents the total mass enclosed on the stellar surface. In order to compare the internal solution with the external, we impose some continuity condition for the metric potentials on the boundary, i.e., $r=R$ as follows
\begin{equation}\label{21}
 {g_{tt}}^{+} = {g_{tt}}^{-}, ~~~~ {g_{rr}}^{+} = {g_{rr}}^{-}, ~~~~ \frac{\partial g_{tt}}{\partial r}^+ =  \frac{\partial g_{tt}}{\partial r}^-.
\end{equation}
Now, by utilizing the above boundary condition, we obtain the unknown constants as
\begin{eqnarray}
\varrho&=& \frac{(R -2M)\bigg(1 -\frac{M}{2\tau M -\tau R +M}\bigg)^{-\tau}}{R},\label{22}\\
h&=&\frac{M}{R^{2} (\tau R -(2\tau +1) M)},\\
o&=&2\tau\bigg(1 -\frac{M}{2\tau M -\tau R +M}\bigg)^{1 -\tau}.\label{24}
\end{eqnarray}
The above-mentioned Eqs. (\ref{22})-(\ref{24}) are very important for further graphical analysis and the physical properties of stellar structure. These equations depend on the constants, $\tau$, $M$, and $R$. Here the numerical values of $o$, $\varrho$, $h$, and $\gamma$ for the considered star models are given below in Table 1-3.

\begin{figure}[h!]
\begin{tabular}{cccc}
\epsfig{file=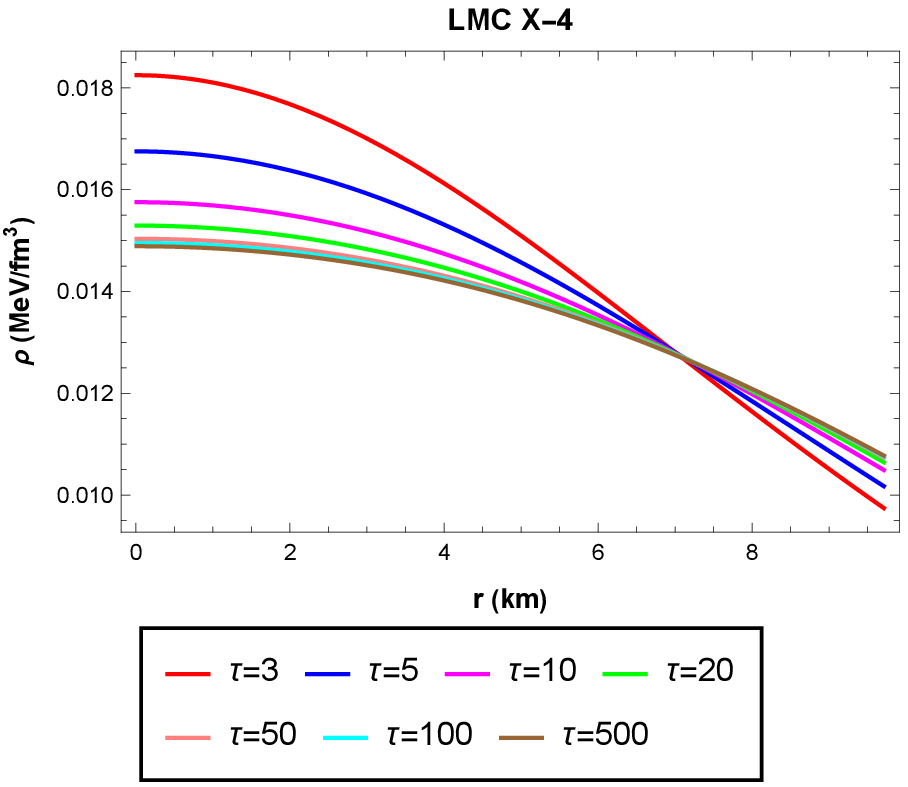,width=5.5cm,height=5.5cm}
\epsfig{file=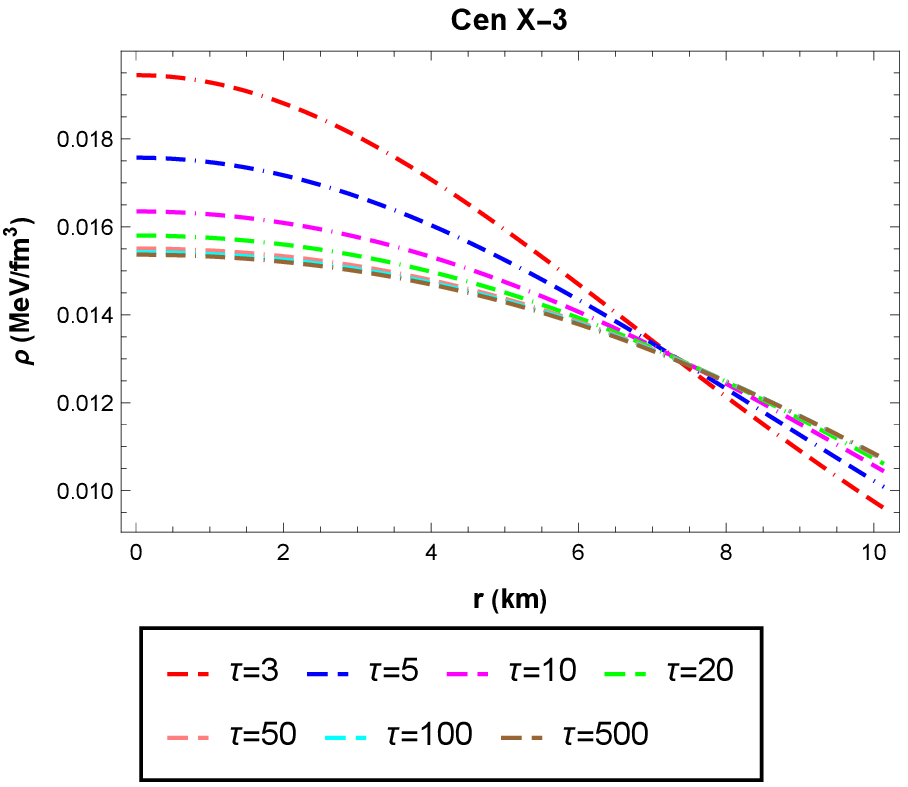,width=5.5cm,height=5.5cm}
\epsfig{file=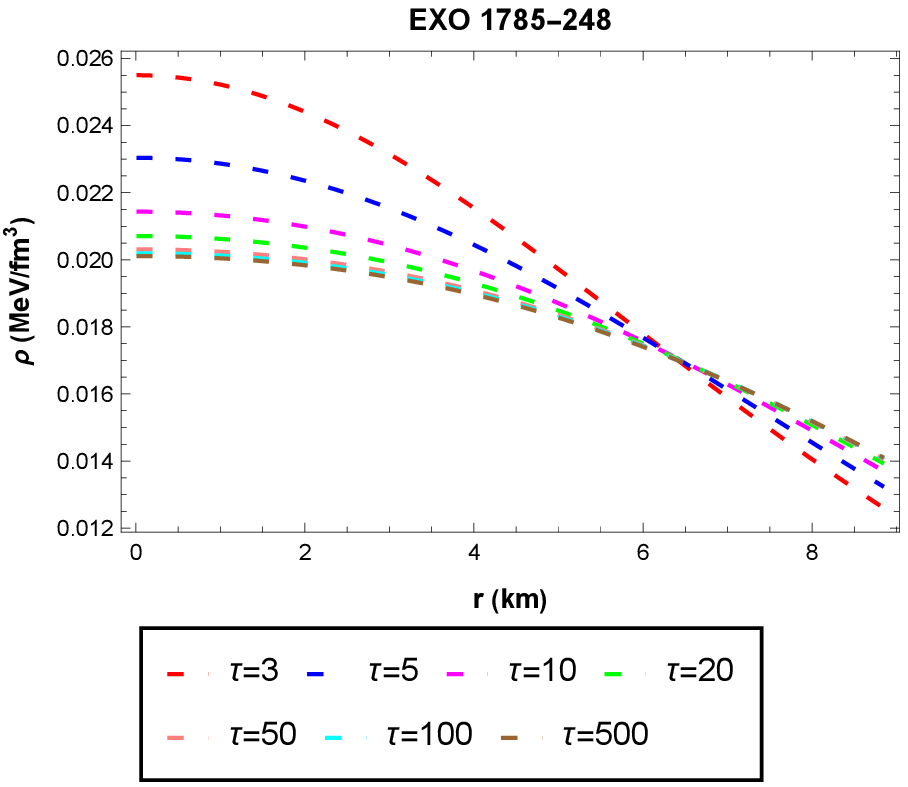,width=5.5cm,height=5.5cm}
\end{tabular}
\caption{\label{Fig.1} Graphical variation of $\rho$ against radial coordinate.}
\end{figure}

\section{Physical features and visual investigation}
In this section, we discuss the physical features and various stellar configurations in the $f(R,\phi,X)$ modified gravity model. Here, we use three different star candidates, i.e., LMC X-4, Cen X-3, and EXO 1785-248. Moreover, to identify the stable configuration in the present study, we utilize the observational data of three compact star candidates, i.e., LMC X-4, Cen X-3, and EXO 1785-248 whose masses are measured by $1.29M/M_{\odot},~1.49M/M_{\odot},~1.30M/M_{\odot}$ with their corresponding radii $9.711~km,~10.136~km, 8.849~km$. These compact stars are X-ray binaries detected by using X-ray telescope, such as NASA's Chandra X-ray observatory or European space Agency's XMM-Newton. Both LMC X-4 and Cen X-3 are categorized as High-Mass X-ray Binaries (HMXB's) due to the nature of their companion stars. In HMXB's, the companion star is typically a massive, young star, often of spectral type O or B. The compact star EXO 1785-248 is also an X-ray binary, but its nature is not well-studied. It is classified as a Low-Mass X-ray Binary (LMXB), which typically consists of a less massive companion star, such as a main-sequence or a compact object, potentially a neutron star or a black hole. Several investigations \cite{abc4,abc5} have been carried out in order to examine the existence of these three compact stars within the framework of general relativity and modified gravity theories. This analysis includes the energy density, pressure components, equation of state parameters, stability analysis, metric potentials, energy condition, adiabatic index and anisotropy parameter. We also investigate some extra features of stellar structuresi.e., mass function, compactness factor and surface redshift. Further, we choose the constants $n=0.1$, $w_0=1\times10^{-5}$, and $\alpha =1\times10^{-6}$ to obtain the desired results. Furthermore, the remaining parameters are already given in tabular form.

\begin{figure}[h!]
\begin{tabular}{cccc}
\epsfig{file=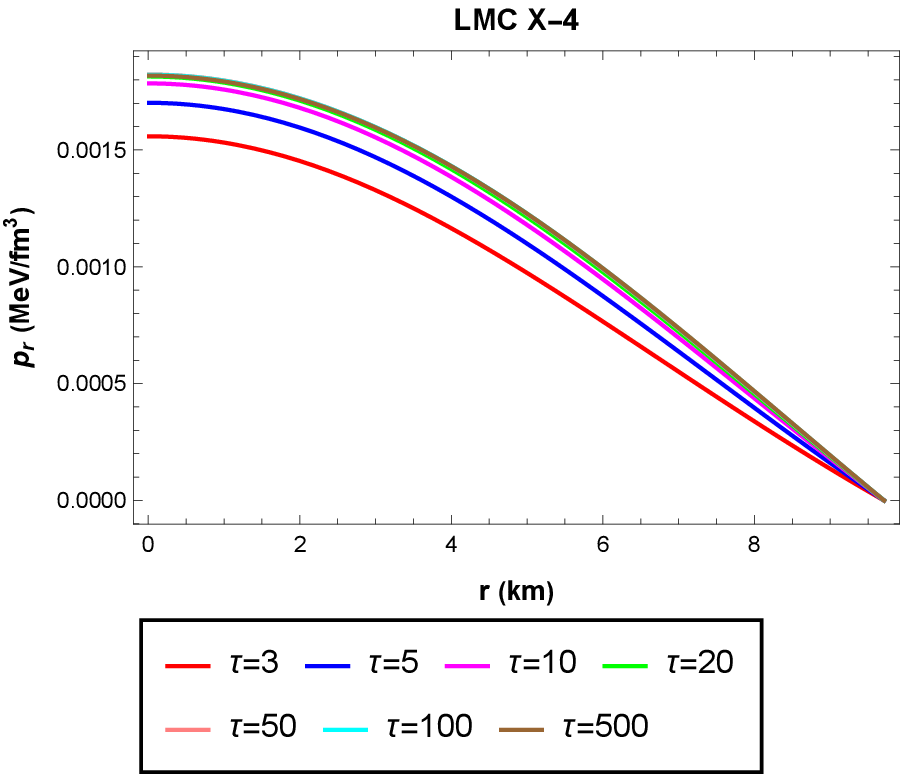,width=5.5cm,height=5.5cm}
\epsfig{file=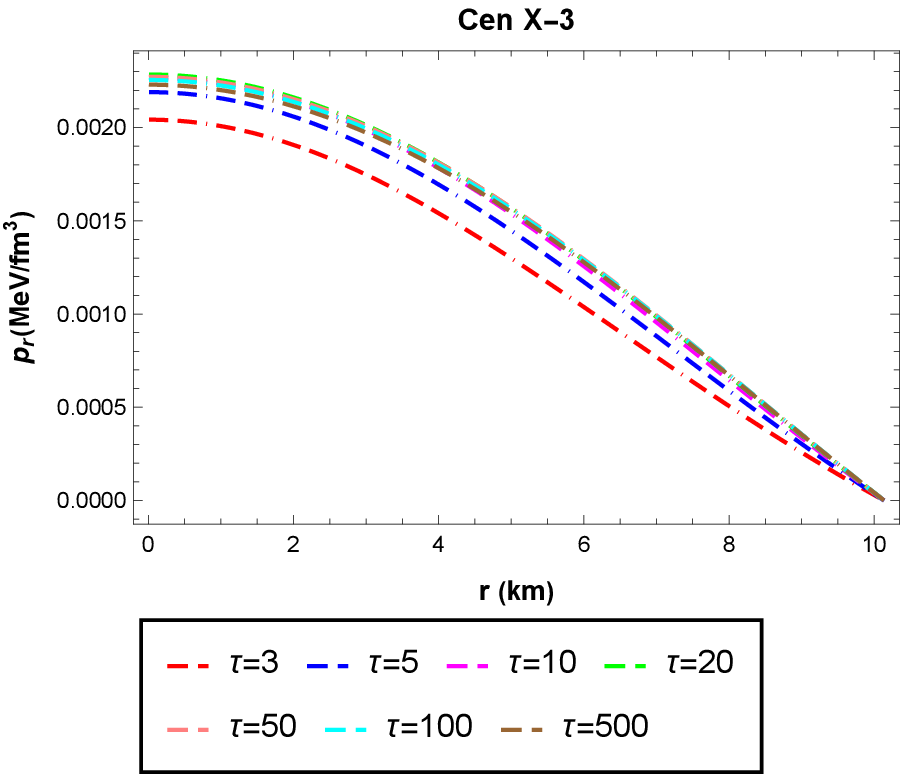,width=5.5cm,height=5.5cm}
\epsfig{file=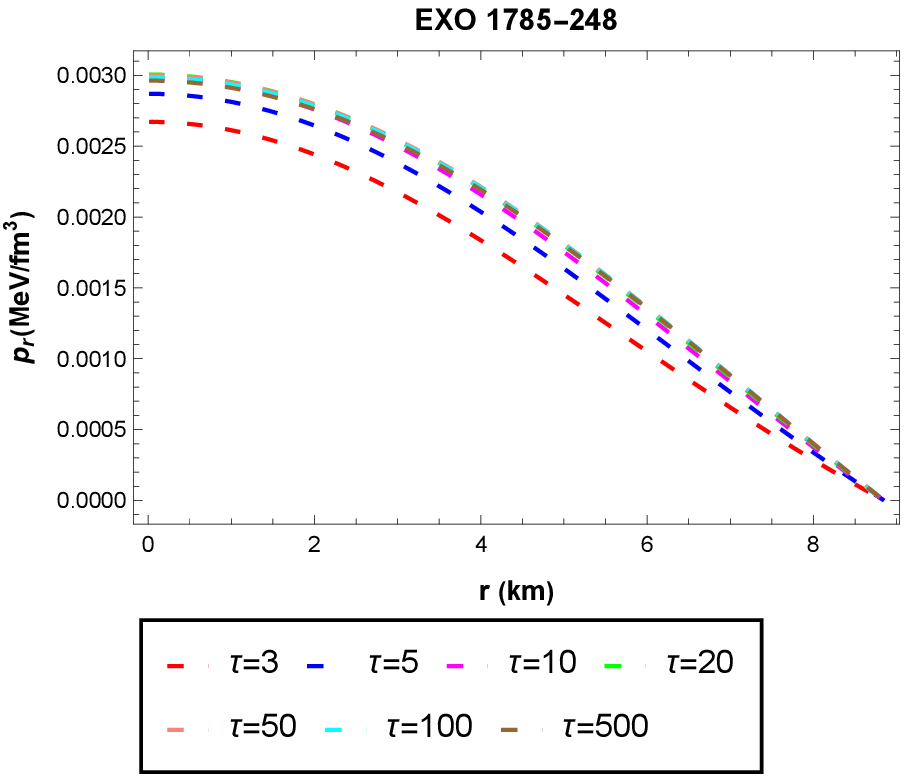,width=5.5cm,height=5.5cm}
\end{tabular}
\caption{\label{Fig.2} Graphical variation of $p_r$ against radial coordinate.}
\end{figure}

\begin{figure}[h!]
\begin{tabular}{cccc}
\epsfig{file=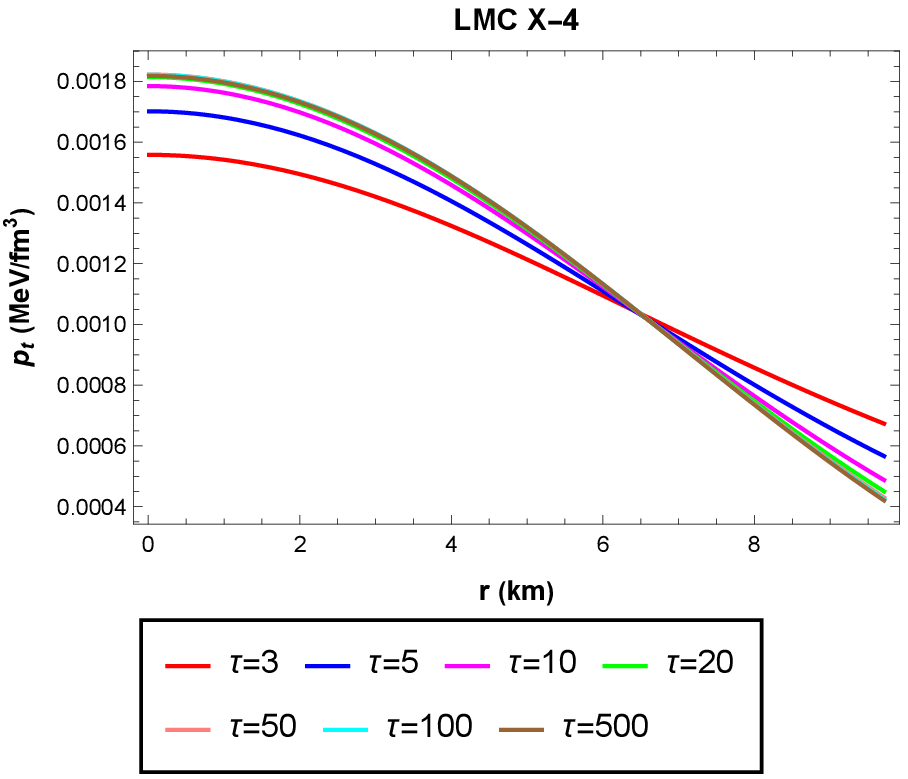,width=5.5cm,height=5.5cm}
\epsfig{file=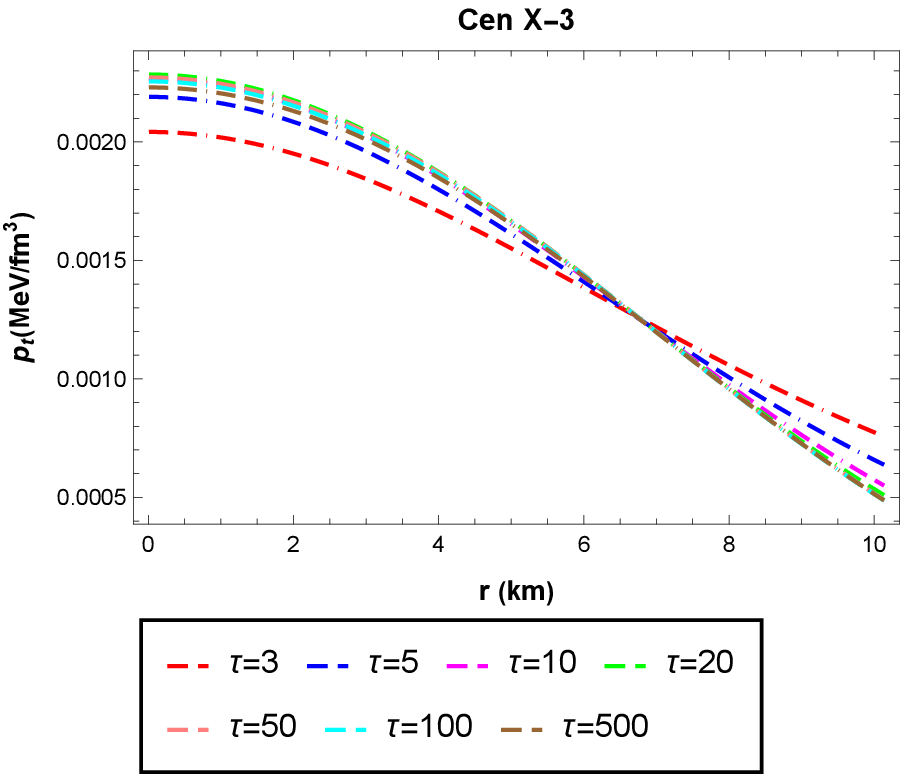,width=5.5cm,height=5.5cm}
\epsfig{file=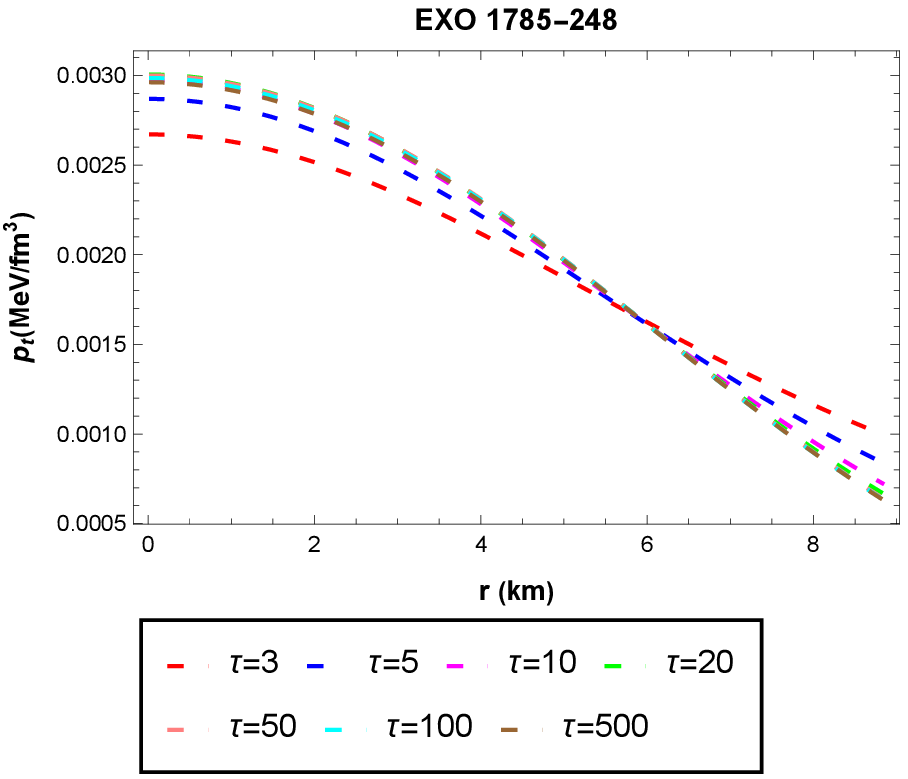,width=5.5cm,height=5.5cm}
\end{tabular}
\caption{\label{Fig.3} Graphical variation of $p_t$ against radial coordinate.}
\end{figure}

\subsection{Energy Density and Pressure Progression}
\begin{figure}[h!]
\begin{tabular}{cccc}
\epsfig{file=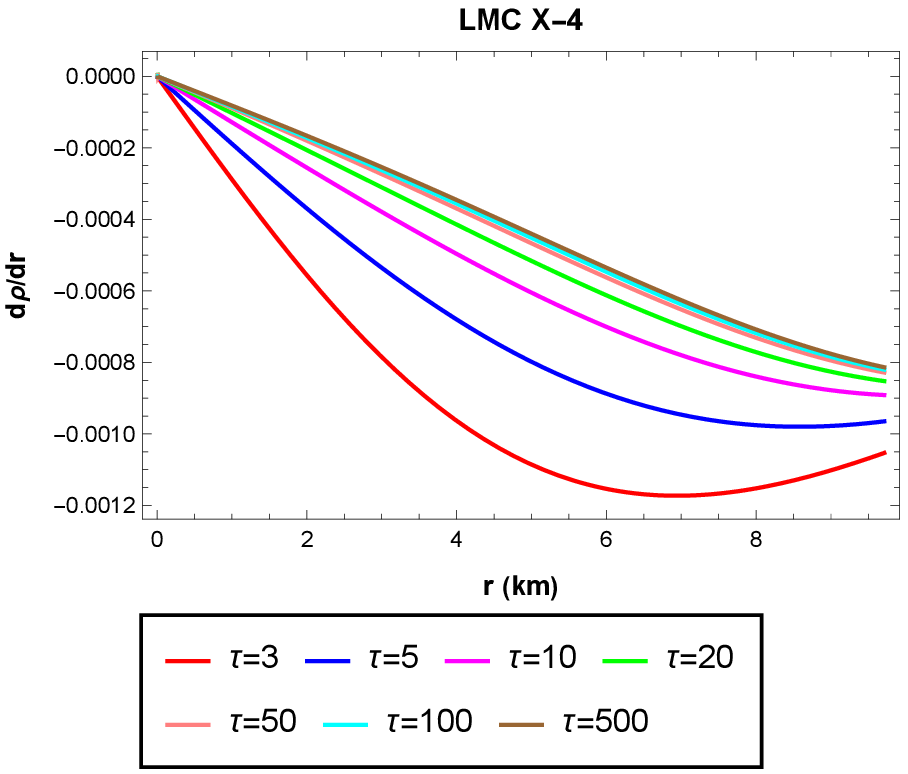,width=5.5cm,height=5.5cm}
\epsfig{file=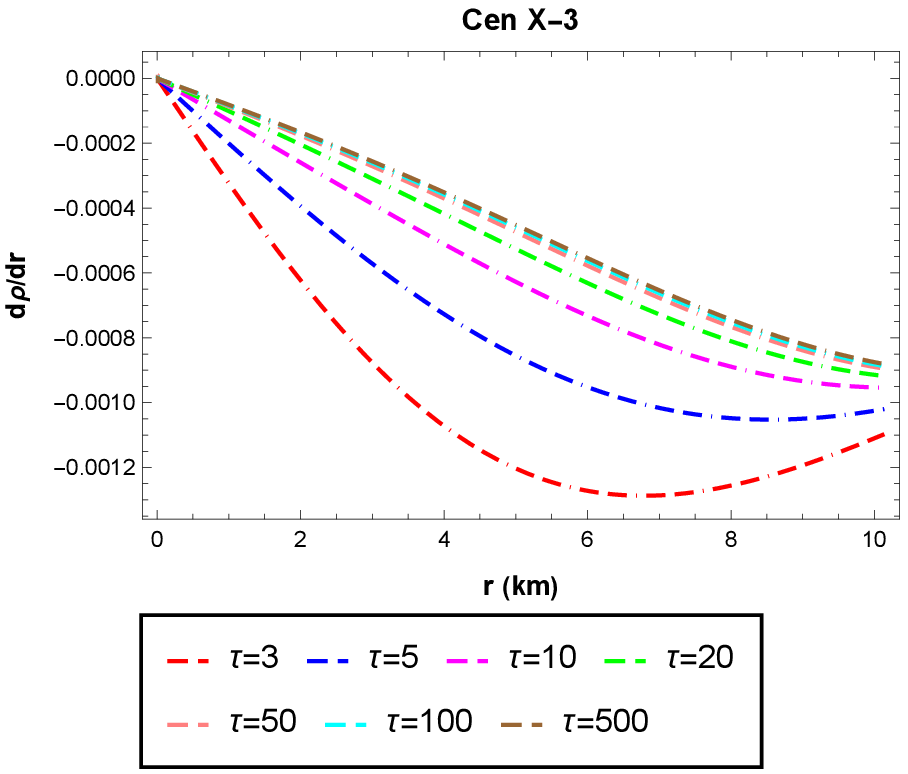,width=5.5cm,height=5.5cm}
\epsfig{file=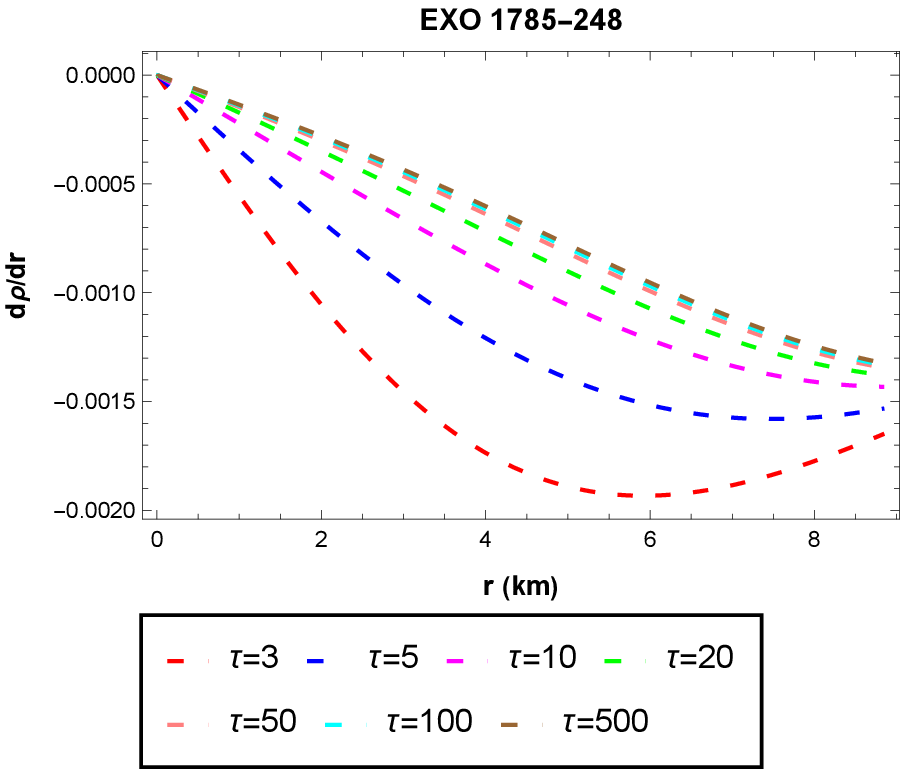,width=5.5cm,height=5.5cm}
\end{tabular}
\caption{\label{Fig.4} Graphical variation of gradient of $\rho$ against radial coordinate.}
\end{figure}

\begin{figure}[h!]
\begin{tabular}{cccc}
\epsfig{file=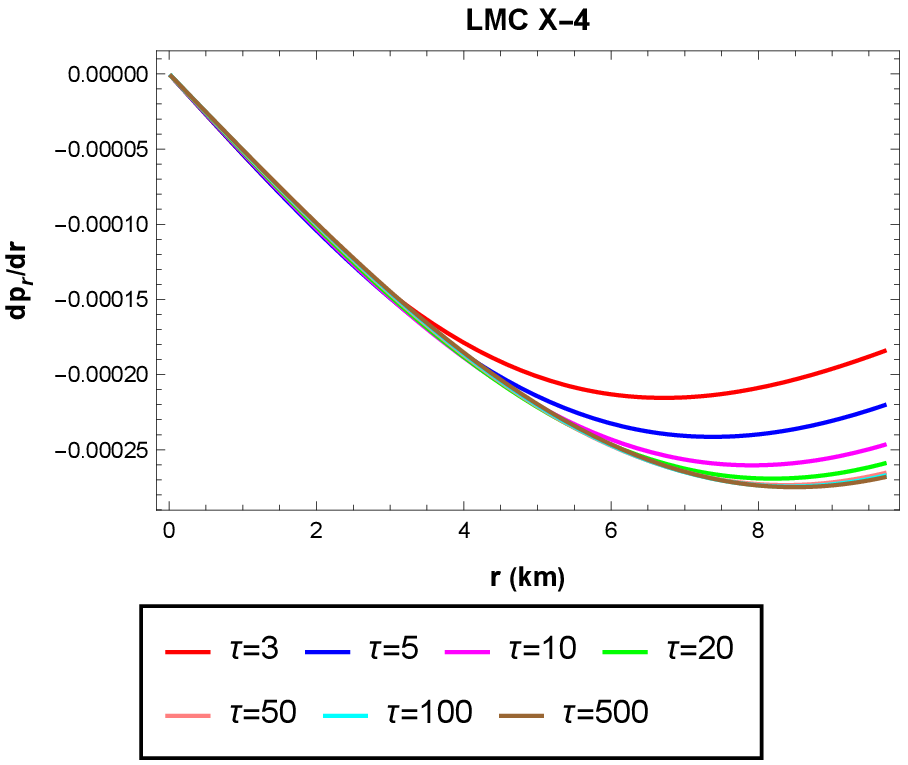,width=5.5cm,height=5.5cm}
\epsfig{file=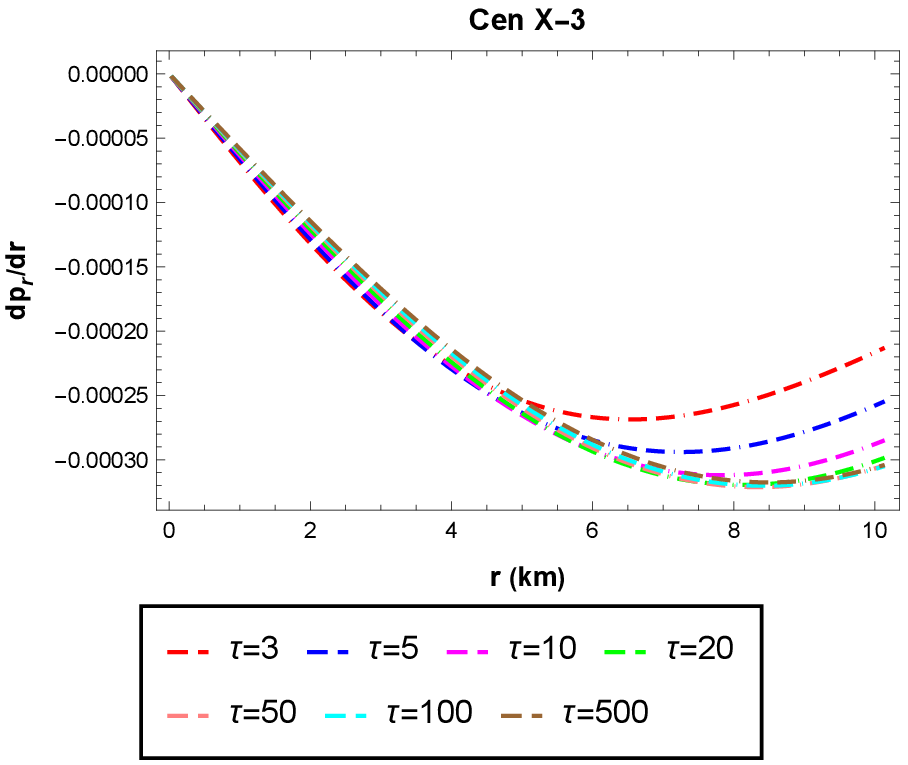,width=5.5cm,height=5.5cm}
\epsfig{file=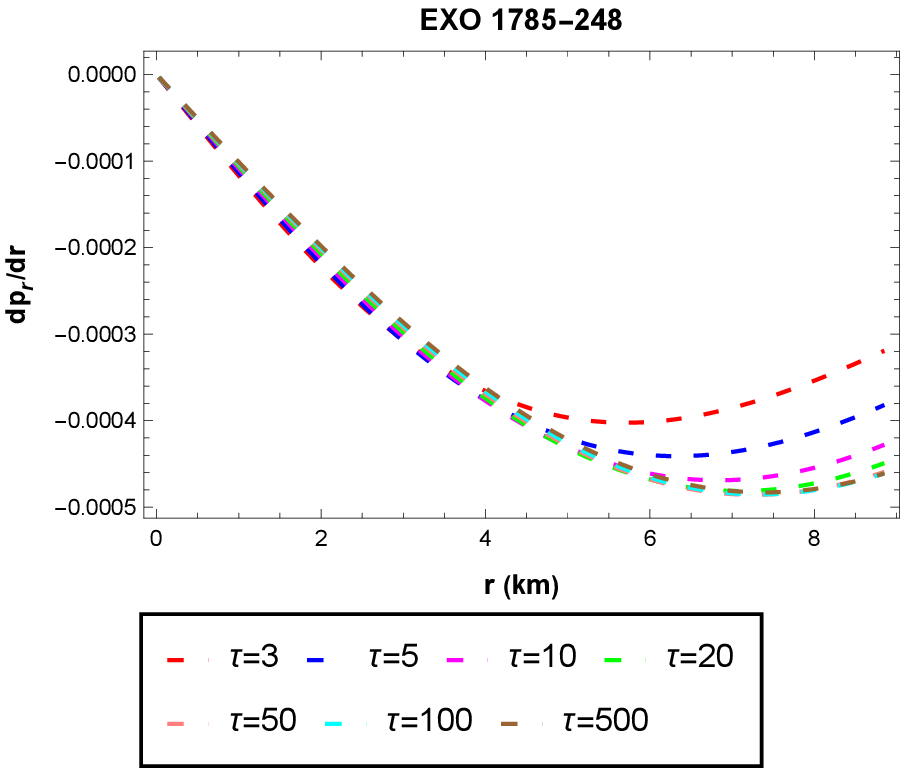,width=5.5cm,height=5.5cm}
\end{tabular}
\caption{\label{Fig.5} Graphical variation of gradient of $p_r$ against radial coordinate.}
\end{figure}

\begin{figure}[h!]
\begin{tabular}{cccc}
\epsfig{file=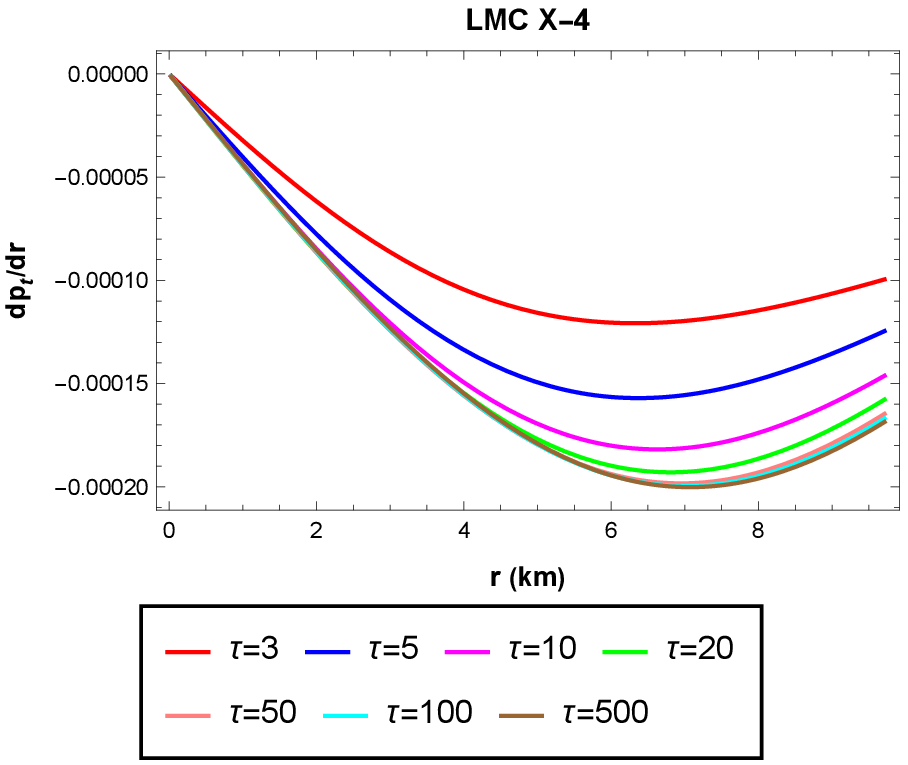,width=5.5cm,height=5.5cm}
\epsfig{file=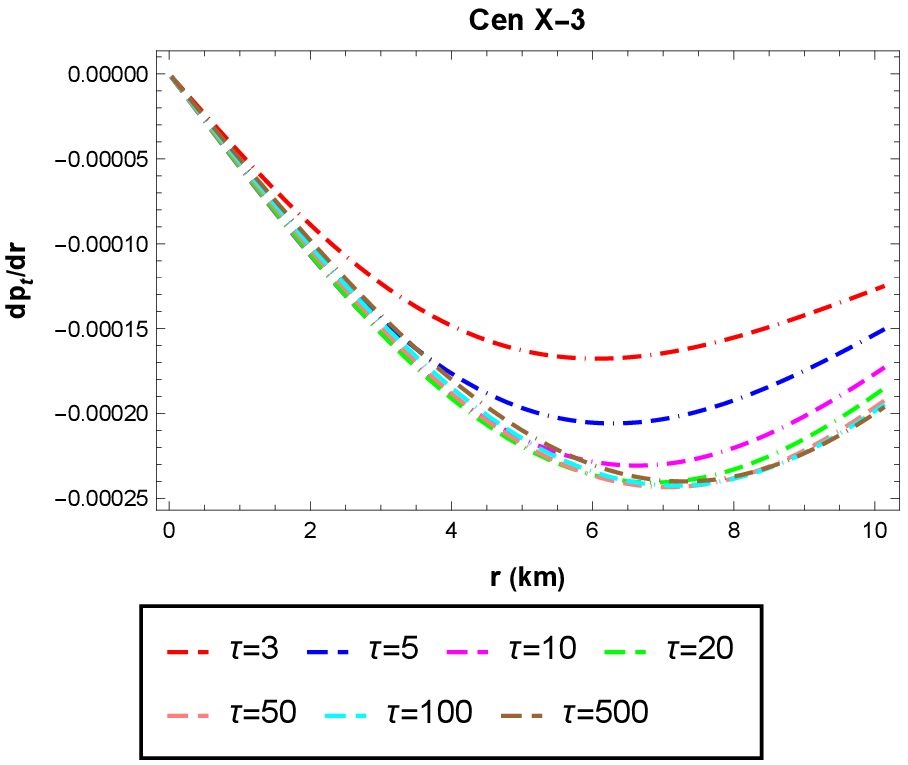,width=5.5cm,height=5.5cm}
\epsfig{file=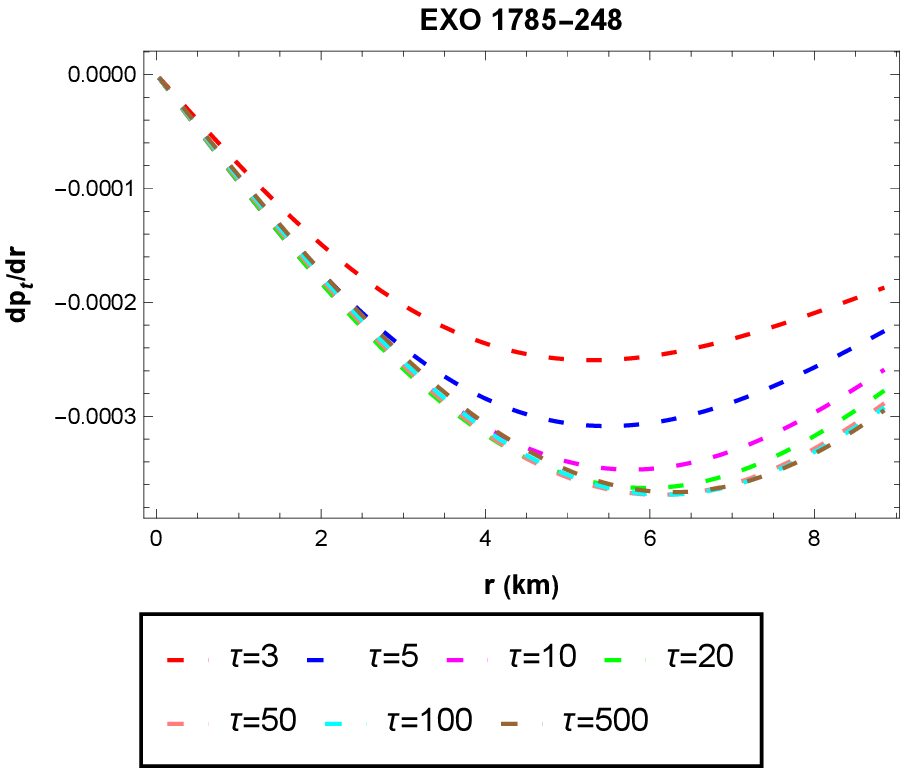,width=5.5cm,height=5.5cm}
\end{tabular}
\caption{\label{Fig.6} Graphical variation of gradient of $p_t$ against radial coordinate.}
\end{figure}

The graphical nature of physical quantities including energy density and pressure components are illustrated in Figs. \ref{Fig.1}-\ref{Fig.3}.
From Fig. \ref{Fig.1}, it can be seen that the graphical analysis of energy density is positive, decreasing and maximum at the core throughout the internal configuration. The graphical analysis of radial pressure is maximum at the center of stellar structure and ultimately vanishes at the boundary of the star as seen in Fig. \ref{Fig.2}. Furthermore, the graphical representation of tangential pressure is positive and decreasing, when we move towards the boundary as seen in Fig. \ref{Fig.3}. The behavior of energy density and pressure components suggest the high compactness of the center of the star, which represent that our model under inspection is feasible for the exterior region of the core. Additionally, we examine the nature of our model by using the first-order derivative test as
\begin{equation}
\frac{d\rho}{dr}<0,~~~~~~\frac{dp_r}{dr}<0,~~~~~~\frac{dp_t}{dr}<0.
\end{equation}
The derivative of energy density, radial pressure and tangential pressure with respect to radial component show negative behavior as seen in Figs. \ref{Fig.4}-\ref{Fig.6}. These gradients must be vanish at the core of the compact star, which is demonstrated as
\begin{equation}
\frac{d\rho}{dr}\mid_{r=0}=0,~~~~~~\frac{dp_r}{dr}\mid_{r=0}=0,~~~~~~\frac{dp_t}{dr}\mid_{r=0}=0.
\end{equation}
It can also be seen from the Figs. \ref{Fig.4}-\ref{Fig.6} that all these components are zero at the center of the considered stellar structures. This is the first derivative test, which demonstrates that all these functions are positive and attained maximum output at the center of the star, which is again a valid condition for the compact star.

\subsection{Anisotropy Evolution}
\begin{figure}[h!]
\begin{tabular}{cccc}
\epsfig{file=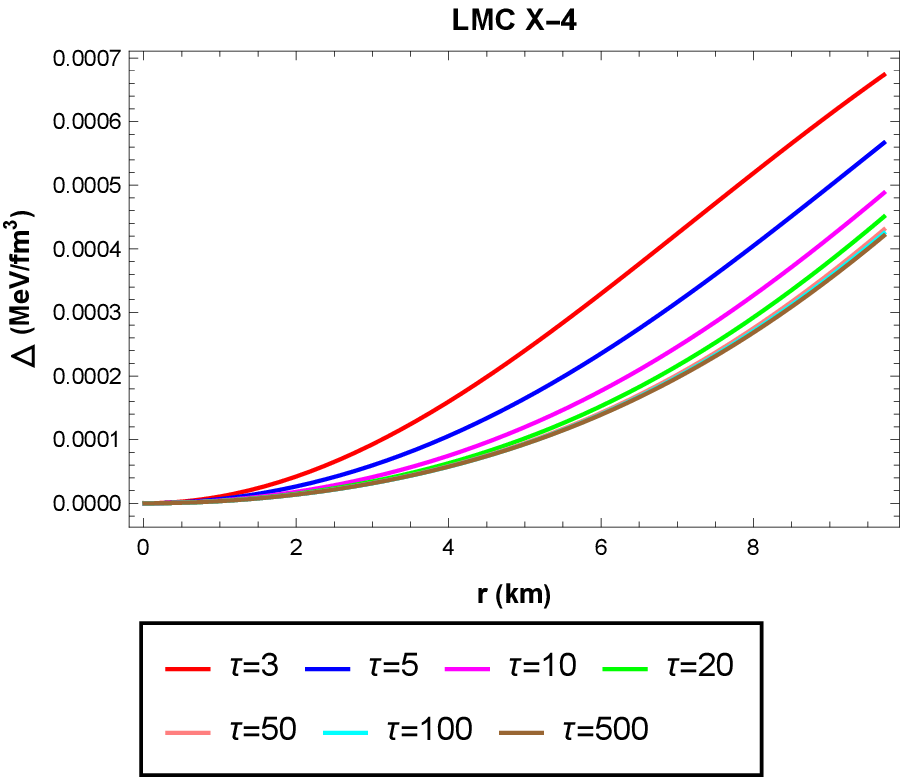,width=5.5cm,height=5.5cm}
\epsfig{file=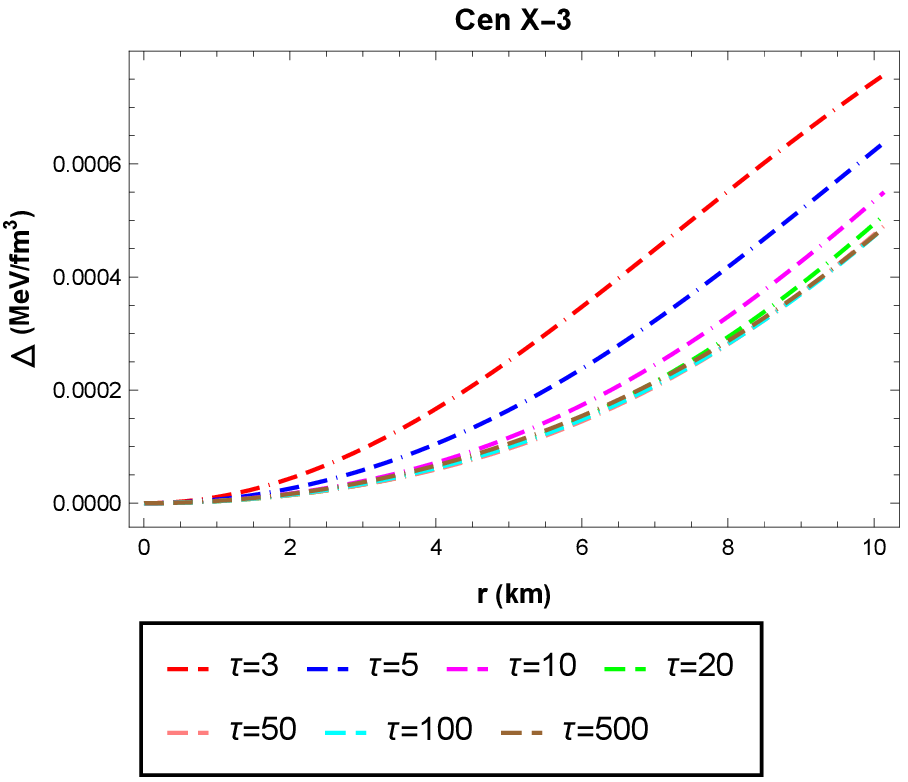,width=5.5cm,height=5.5cm}
\epsfig{file=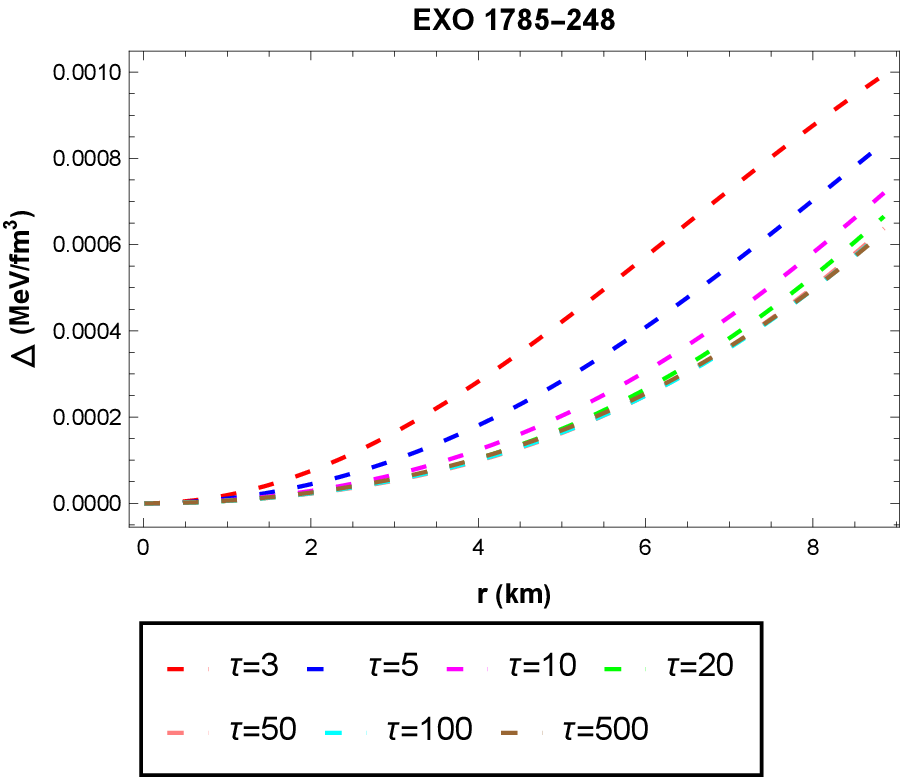,width=5.5cm,height=5.5cm}
\end{tabular}
\caption{\label{Fig.7} Graphical variation of anisotropy $(\Delta)$ against radial coordinate.}
\end{figure}

Anisotropy is a well-known parameter that depicts the information regarding anisotropic behavior of stellar configuration. The difference of tangential pressure and radial component (i.e.,  $p_t - p_r$) is known as anisotropic function, which is considered a very attractive feature to study the compact star. Anisotropy is denoted by $\Delta$ and denoted by
\begin{equation}
 \Delta = p_t - p_r.
\end{equation}
There is an interesting fact that if $\Delta=0$, then there is isotropic pressure in the matter distribution. If anisotropic measurement is positive $\Delta>0$, then anisotropic force is outward. On the other hand, force is inward if an anisotropic measurement is negative ($\Delta<0$) \cite{ad15}. In our current manuscript, the anisotropy is positive, directed outward and zero at the core of the considered stars as seen in Fig. \ref{Fig.7}.

\subsection{Equations of State Parameters}
In literature, there are several equations of state (EoS) parameters but we prefer radial EoS parameter $(EoS)_r$ and tangential EoS parameter $(EoS)_t$ \cite{ad16}, which is defined as
\begin{equation}
(EoS)_{r} = \frac{p_{r}}{\rho},~~~~~~~~~
(EoS)_{t} = \frac{p_{t}}{\rho}.
\end{equation}
\begin{figure}[h!]
\begin{tabular}{cccc}
\epsfig{file=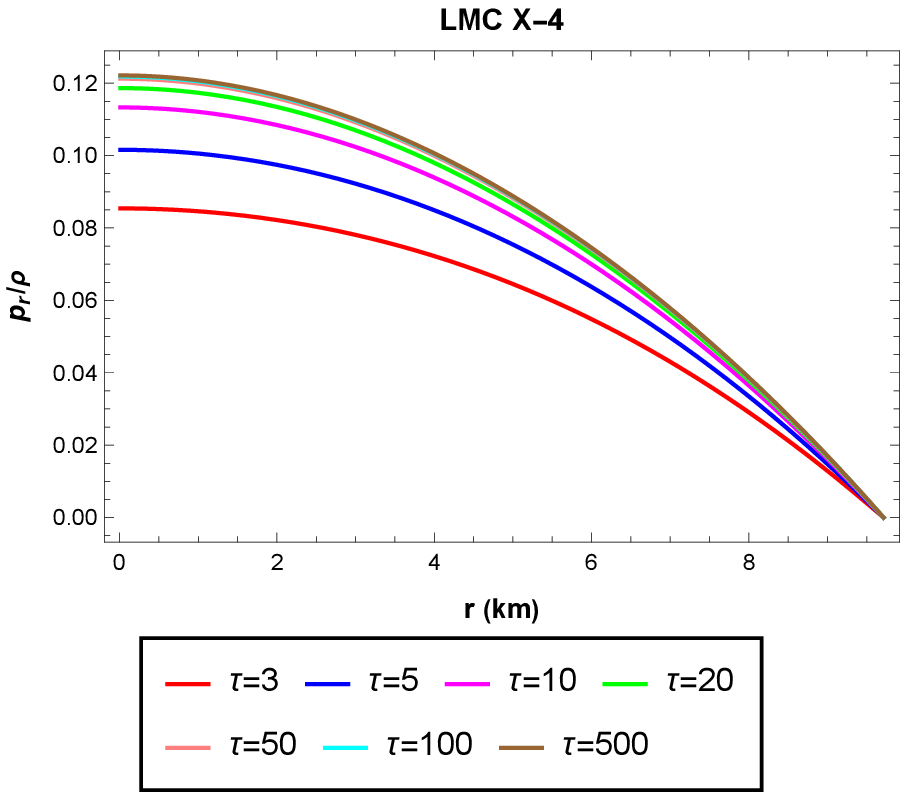,width=5.5cm,height=5.5cm}
\epsfig{file=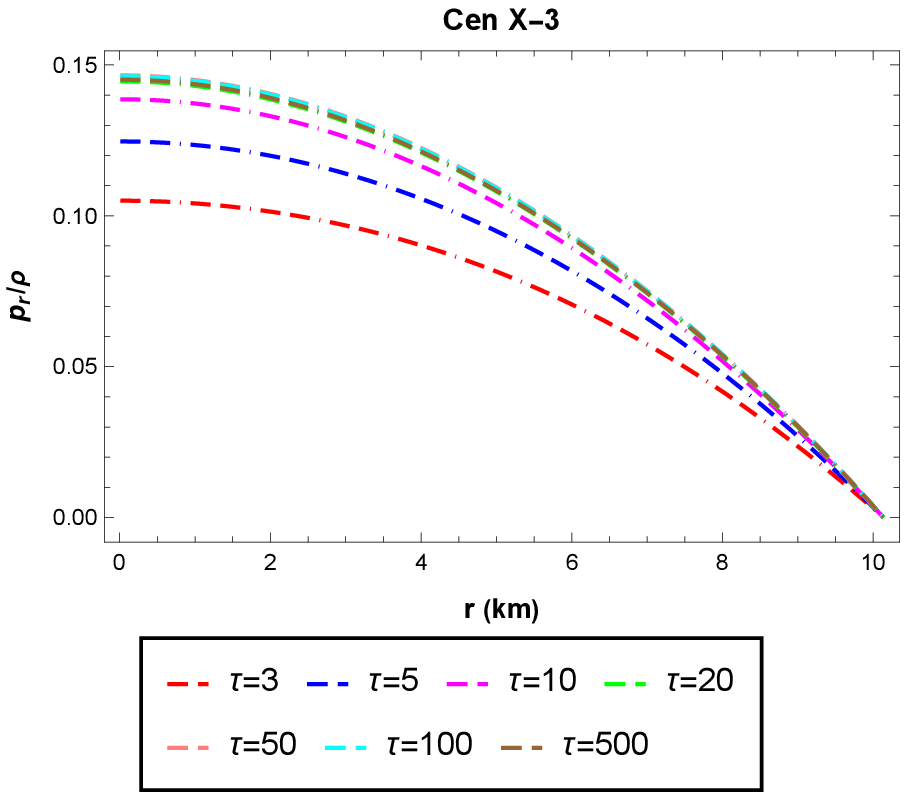,width=5.5cm,height=5.5cm}
\epsfig{file=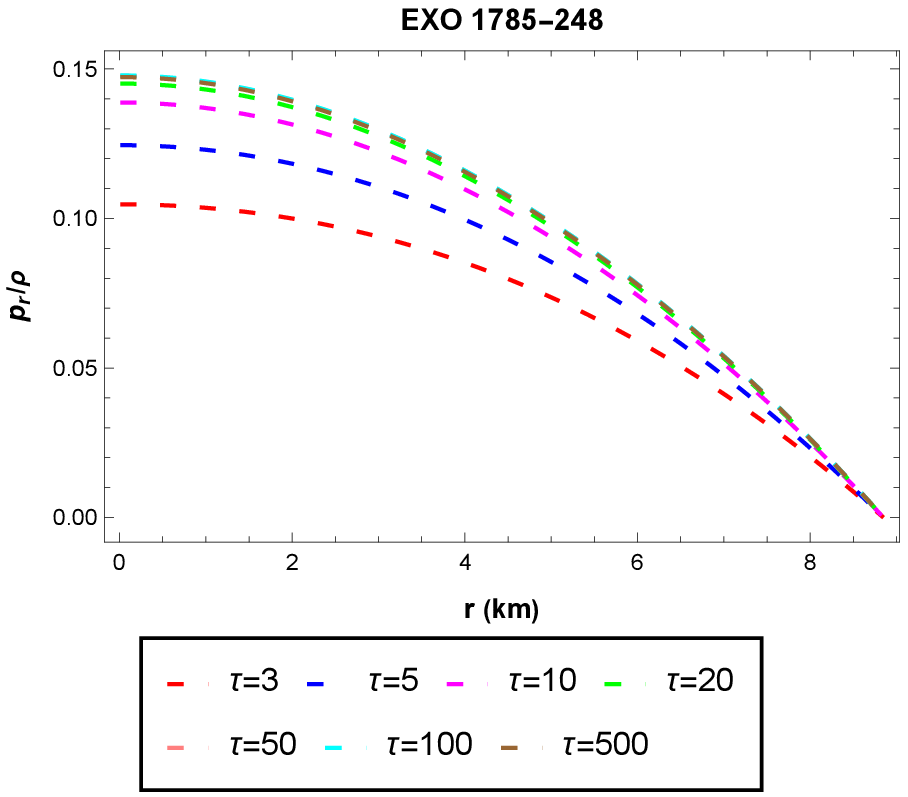,width=5.5cm,height=5.5cm}
\end{tabular}
\caption{\label{Fig.8} Graphical variation of $(EoS)_r$ parameter against radial coordinate.}
\end{figure}

\begin{figure}[h!]
\begin{tabular}{cccc}
\epsfig{file=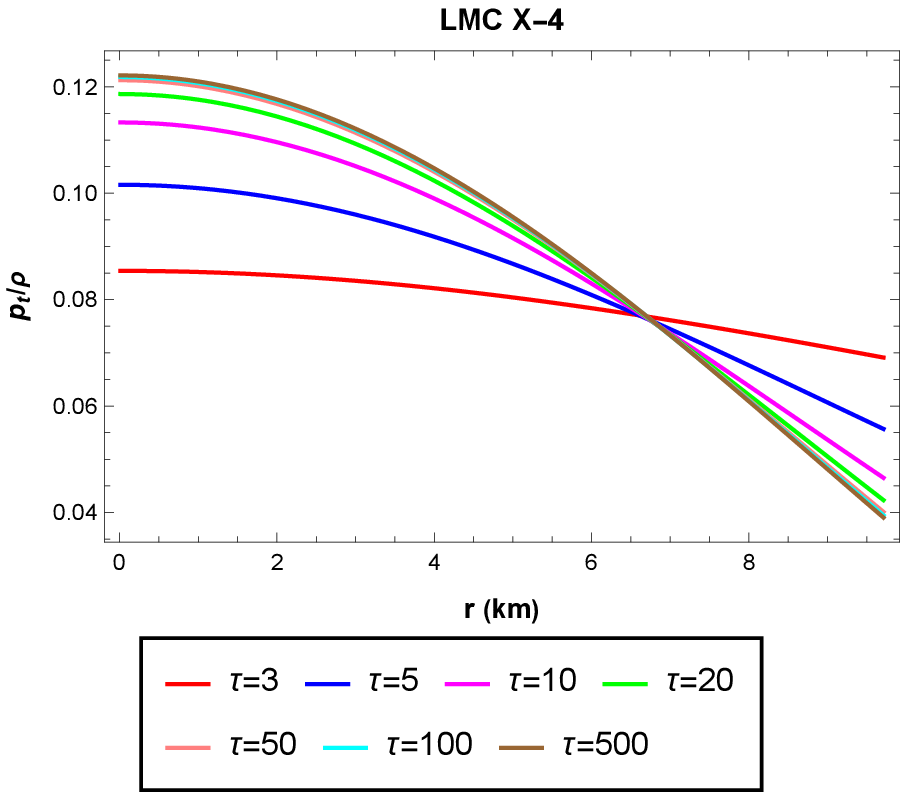,width=5.5cm,height=5.5cm}
\epsfig{file=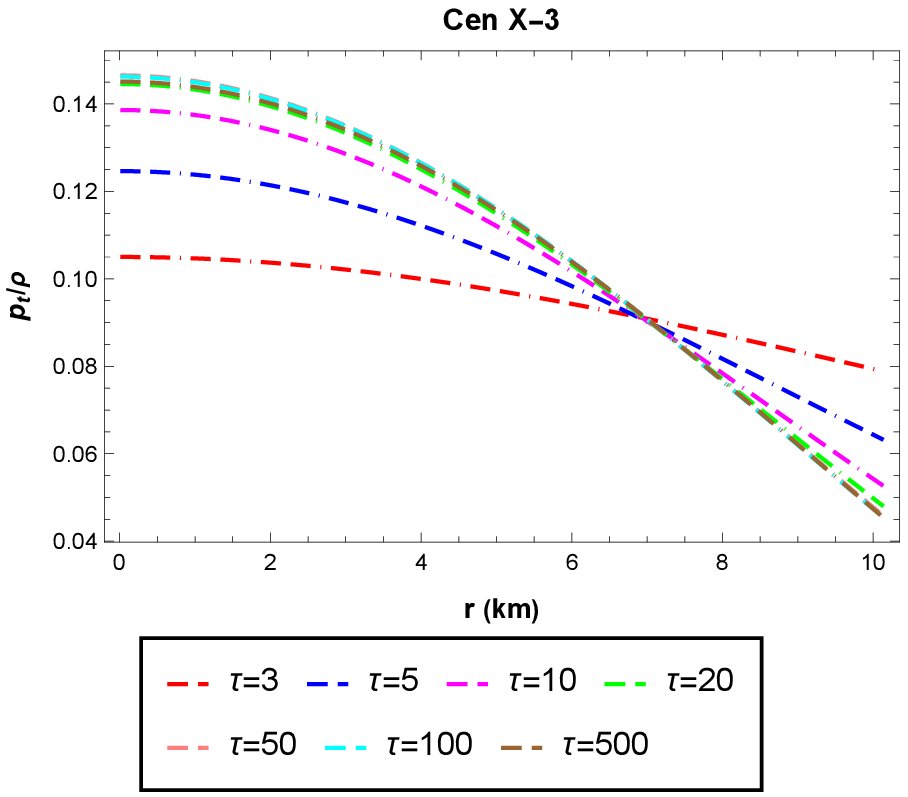,width=5.5cm,height=5.5cm}
\epsfig{file=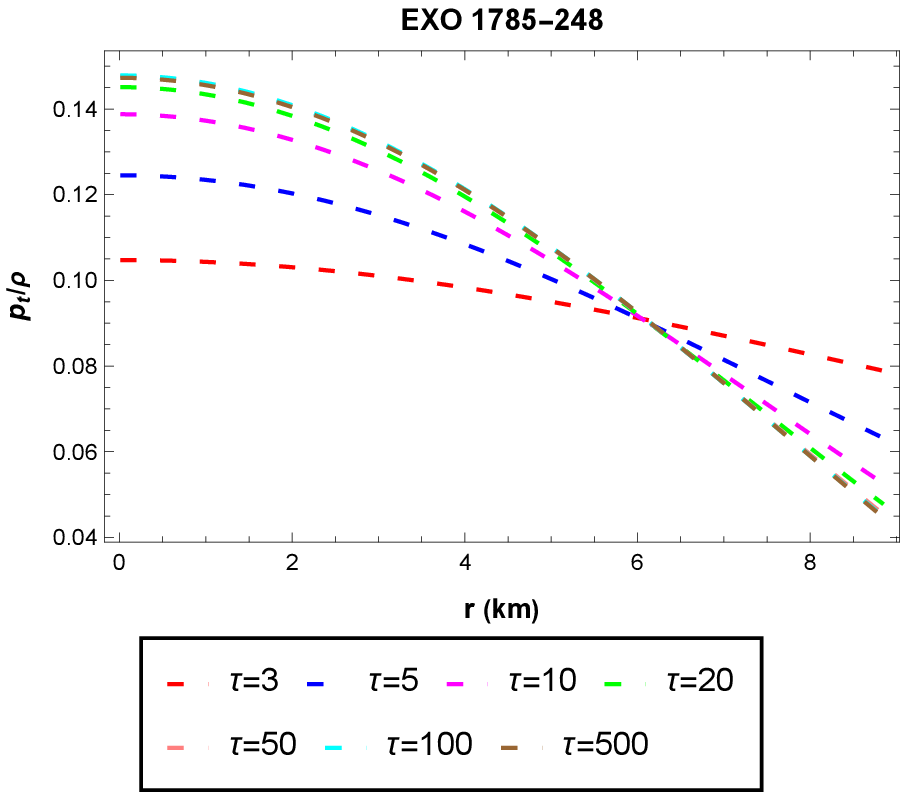,width=5.5cm,height=5.5cm}
\end{tabular}
\caption{\label{Fig.9} Graphical variation of $(EoS)_t$ parameter against radial coordinate.}
\end{figure}

In stellar configuration, the EoS parameters plays an important role in determining the hydrostatic equilibrium which refers to the balance between inward gravitational force and the outward pressure force within a star. Furthermore, the mandatory and adequate condition for these two parameters is that the values of $(EoS)_r$ and $(EoS)_t$ lies between $0$ and $1$. If this parameter is too high (greater than 1), it can lead to an unstable configuration and gravitational collapse. Similarly, if EoS parameter is too low (less than 0), it can also lead to instability and expansion of the stellar configuration. Moreover, the value of EoS parameter between 0 and 1 is crucial for compact star stability because it implies the presence of degeneracy pressure, which counteracts gravitational collapse, preventing the star from further compression. The EoS value in this range signifies the existence of stable white dwarfs, neutron stars, and other compact objects.
It can be observed from Figs. \ref{Fig.8} and \ref{Fig.9} that the behavior of radial and tangential EoS parameters is maximum at core of star, monotonically decreasing towards the boundary, and lying between 0 and 1. From these figures, we note that the tangential pressure and radial pressure are less than the energy density throughout the internal configuration.

\subsection{Gravitational Metric Potential}
\begin{figure}[h!]
\begin{tabular}{cccc}
\epsfig{file=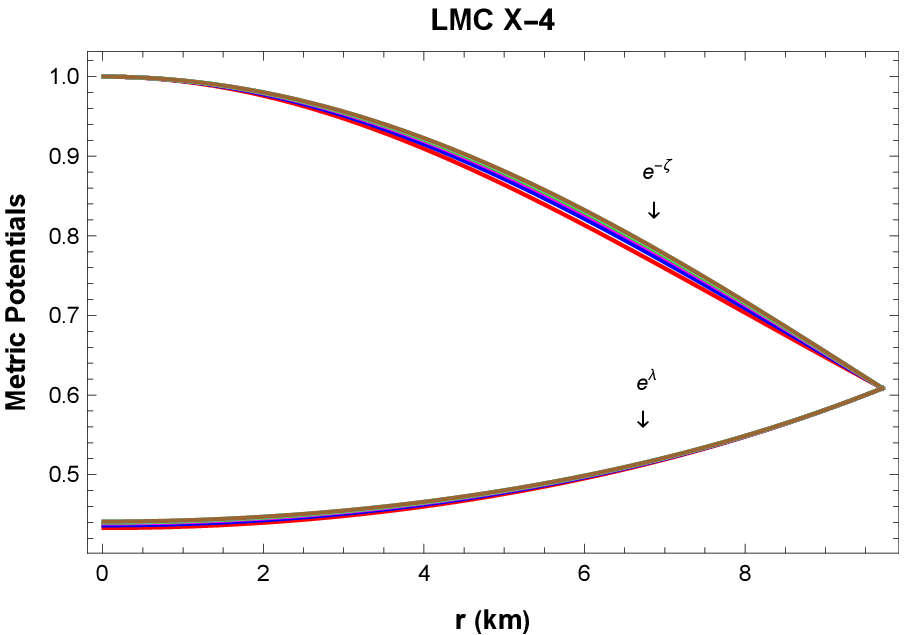,width=5.5cm,height=5.5cm}
\epsfig{file=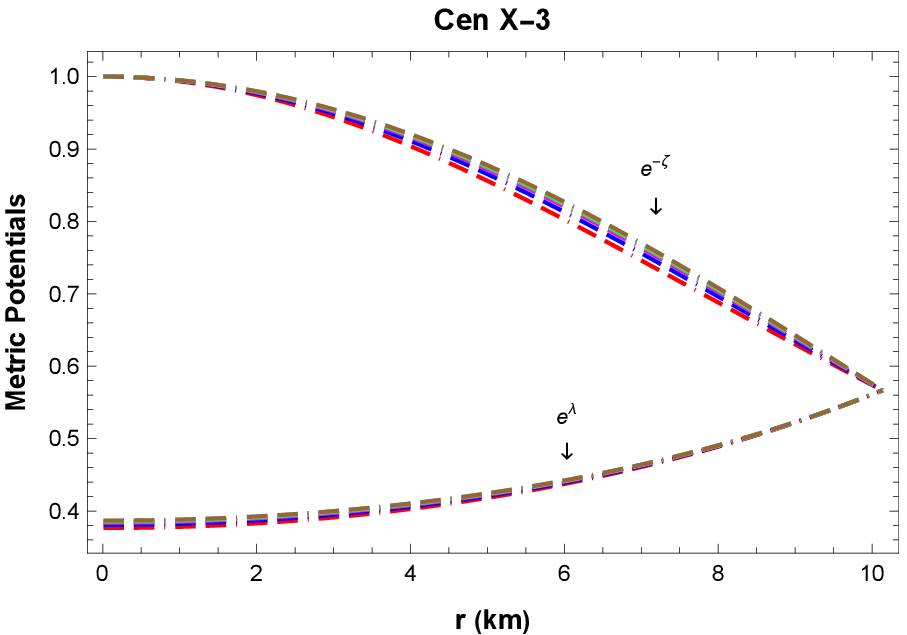,width=5.5cm,height=5.5cm}
\epsfig{file=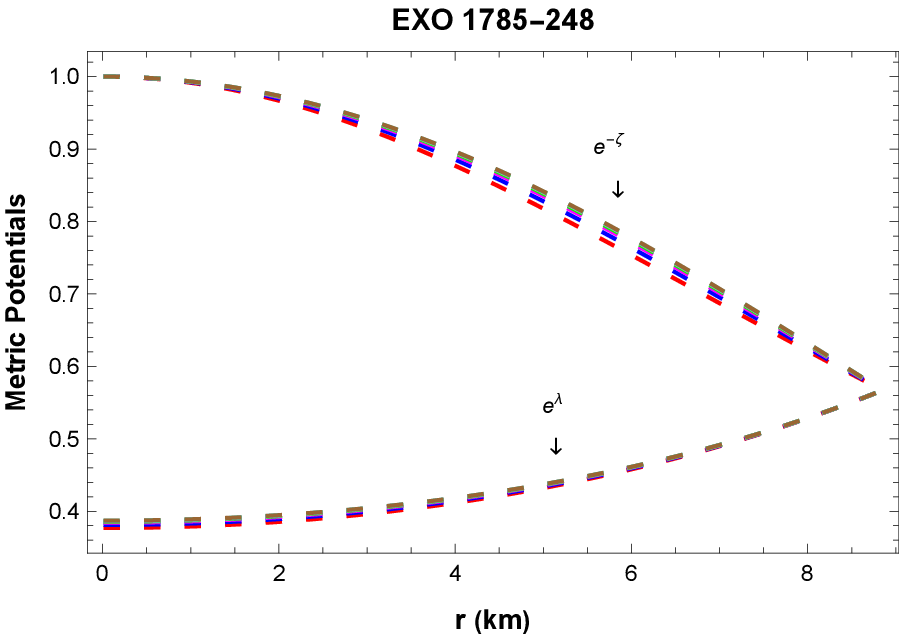,width=5.5cm,height=5.5cm}
\end{tabular}
\caption{\label{Fig.10} Graphical representation of metric potentials against radial coordinate.}
\end{figure}

In the study of stellar spheres, the presence of geometric singularity is always a crucial feature. We investigate the compulsory condition for the metric potential, which is $e^{\zeta(r=0)}=1$ and $e^{\lambda(r=0)}\neq0$. From Fig. \ref{Fig.10}, it can be observed that the behavior of metric potentials is finite, positive at the center, and increasing towards the boundary, which shows that the present stellar structures are free from singularity. Moreover, the matching of the metric potentials at the boundary of the star is a consequence of the requirement of continuity and smoothness in the gravitational field. At the boundary of the star, the interior solution describing the gravitational field must smoothly match with exterior solution, which corresponds to the gravitational field outside the star.

\subsection{Stability Analysis}
\begin{figure}[h!]
\begin{tabular}{cccc}
\epsfig{file=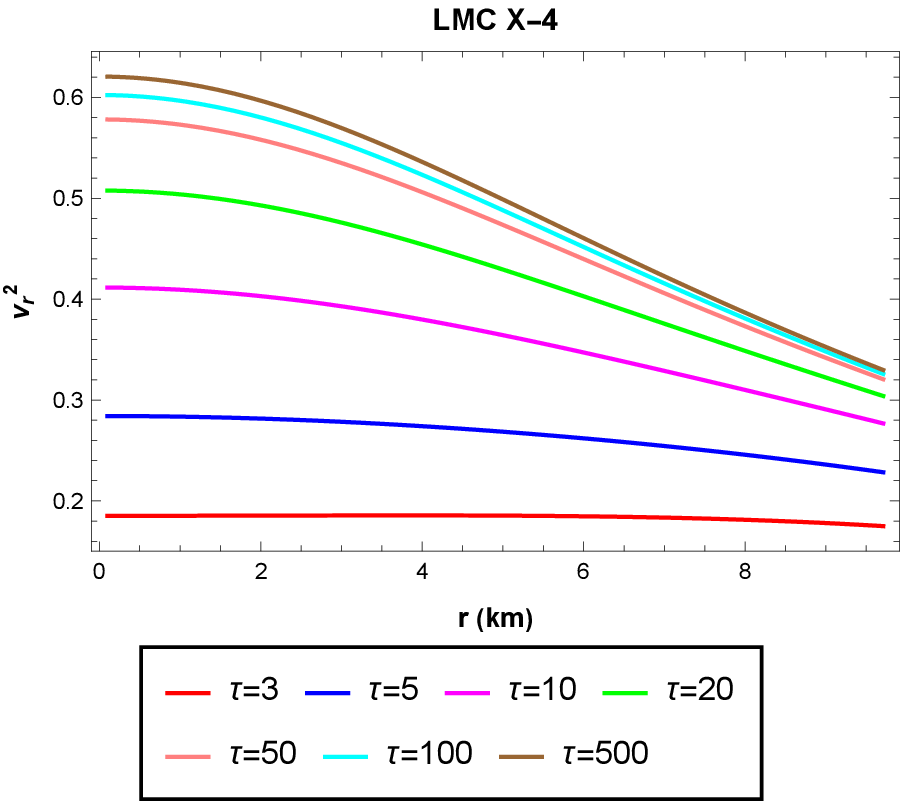,width=5.5cm,height=5.5cm}
\epsfig{file=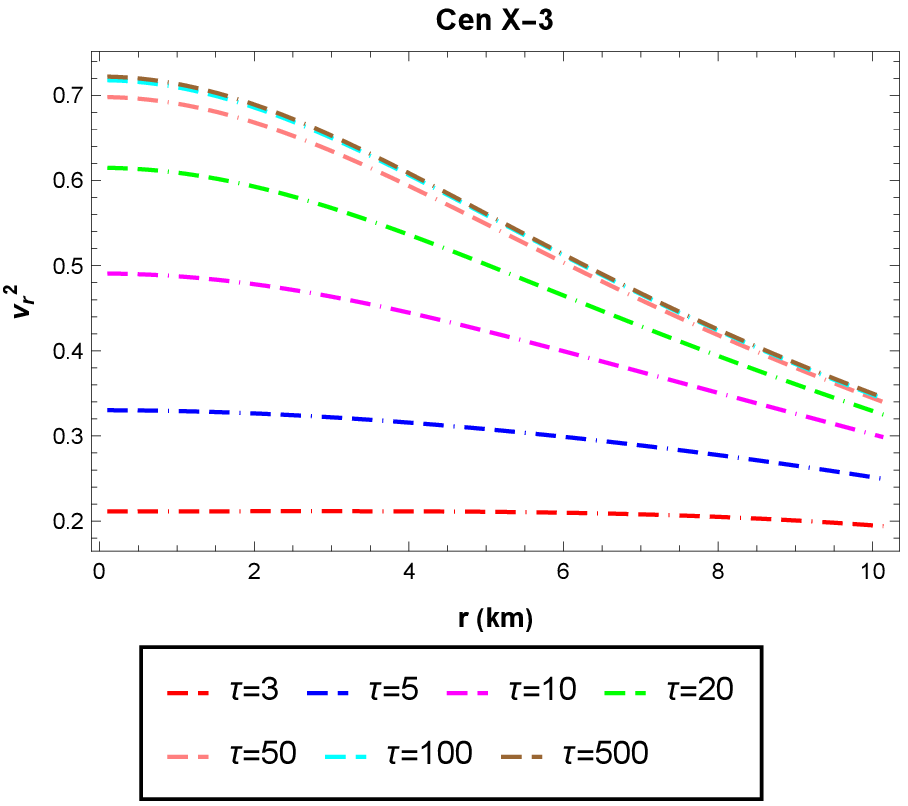,width=5.5cm,height=5.5cm}
\epsfig{file=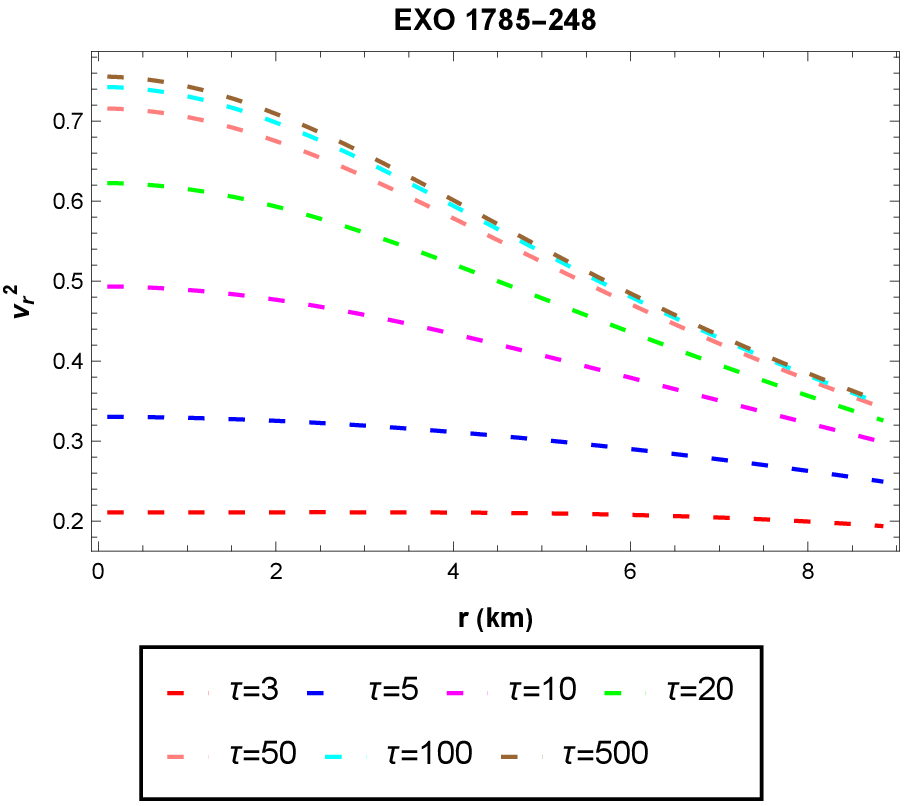,width=5.5cm,height=5.5cm}
\end{tabular}
\caption{\label{Fig.11} Graphical representation of $\nu^{2}_{sr}$ against radial coordinate.}
\end{figure}

\begin{figure}[h!]
\begin{tabular}{cccc}
\epsfig{file=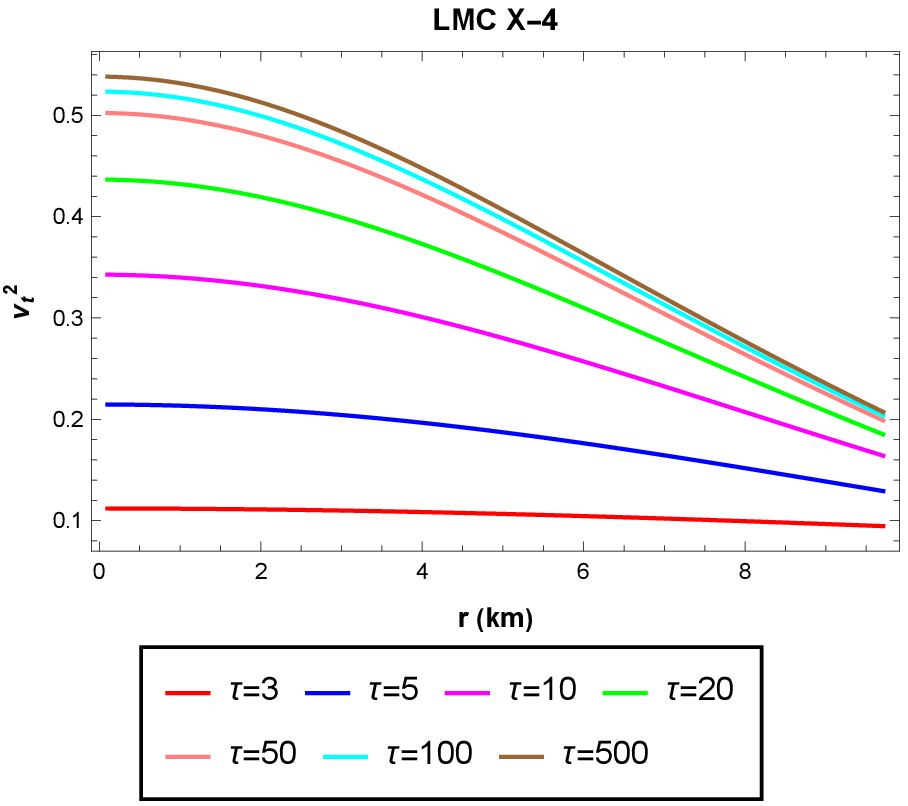,width=5.5cm,height=5.5cm}
\epsfig{file=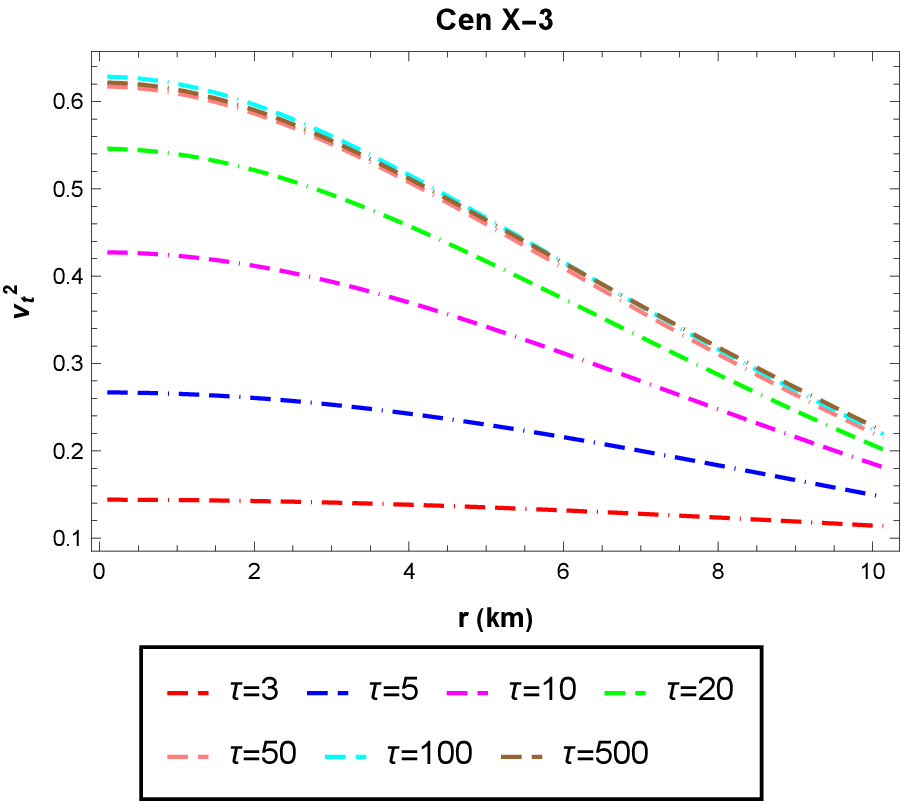,width=5.5cm,height=5.5cm}
\epsfig{file=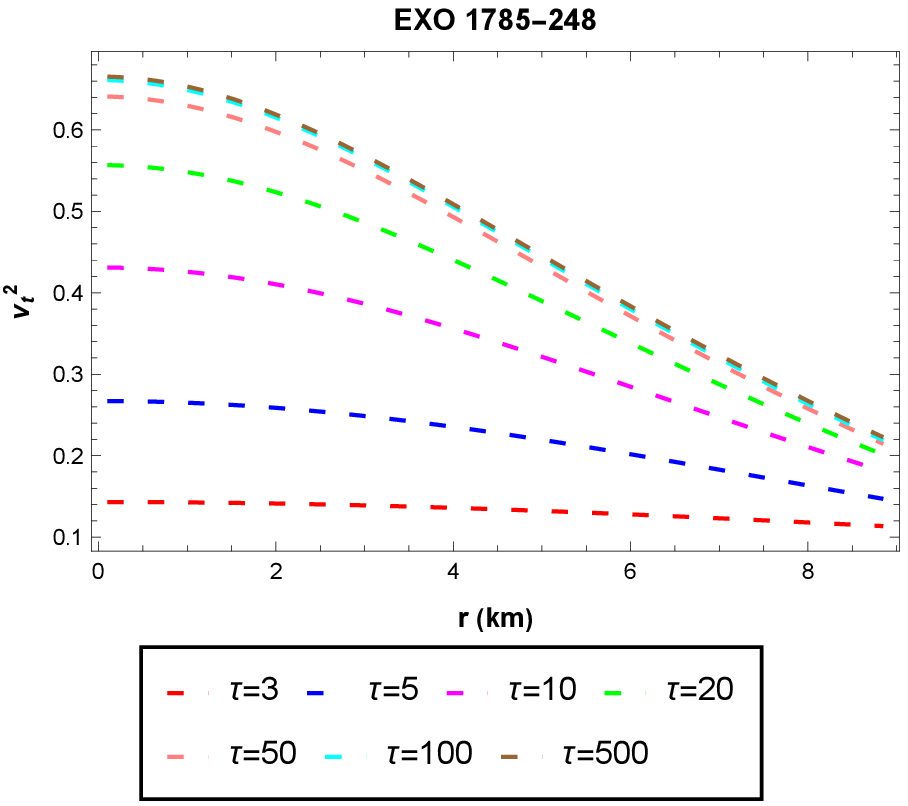,width=5.5cm,height=5.5cm}
\end{tabular}
\caption{\label{Fig.12} Graphical representation of $\nu^{2}_{st}$ against radial coordinate. }
\end{figure}

\begin{figure}[h!]
\begin{tabular}{cccc}
\epsfig{file=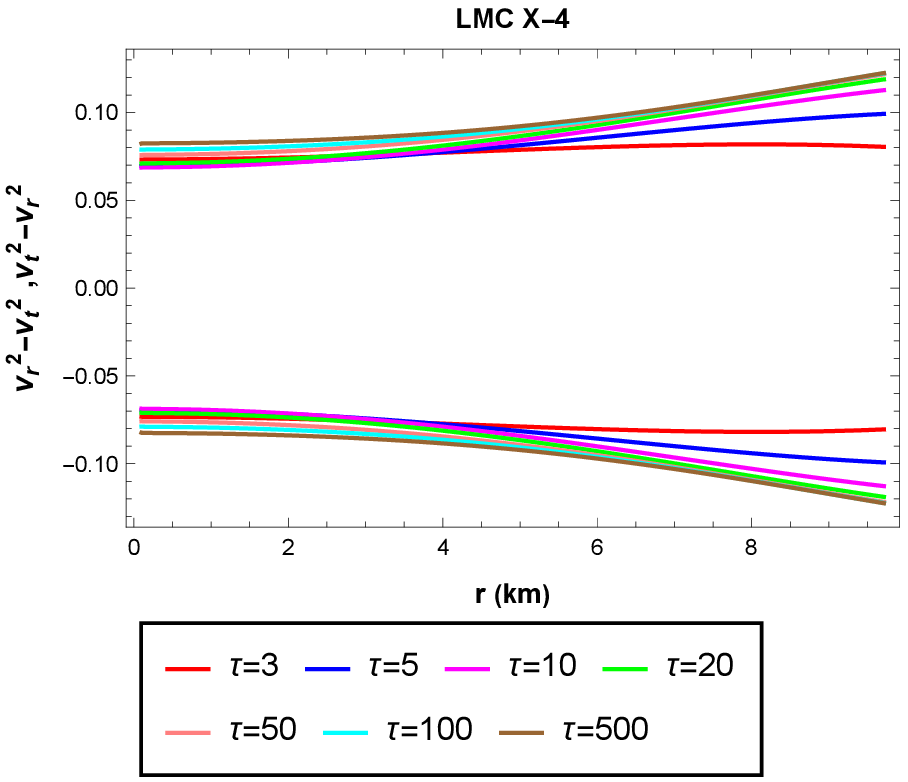,width=5.5cm,height=5.5cm}
\epsfig{file=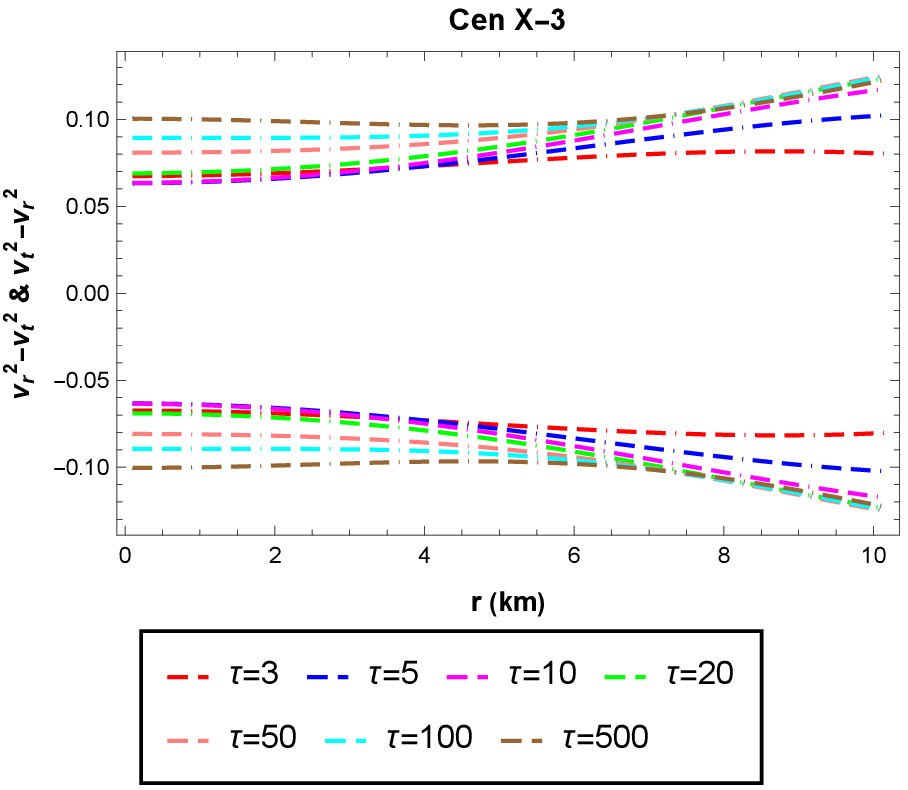,width=5.5cm,height=5.5cm}
\epsfig{file=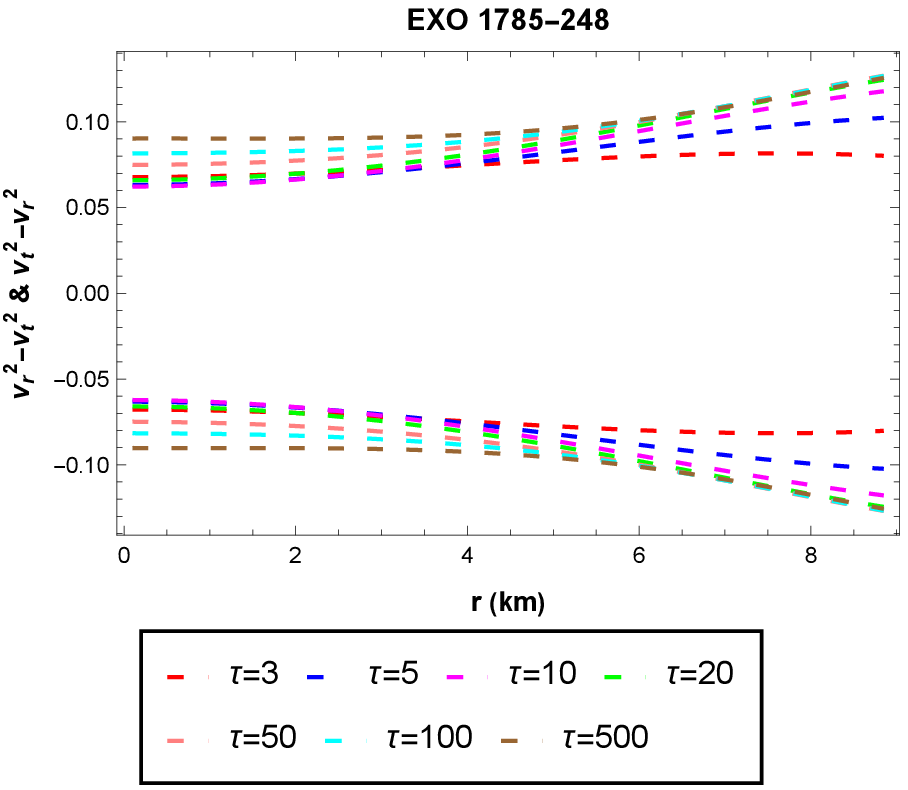,width=5.5cm,height=5.5cm}
\end{tabular}
\caption{\label{Fig.13} Graphical representation of $-1\leq \nu^{2}_{t} - \nu^{2}_{r} \leq 0$ and $0 \leq \nu^{2}_{r} - \nu^{2}_{t} \leq 1$.}
\end{figure}

To examine the stability of compact stars within the context of modified gravity is an important part of stellar strucuture study Pretel et al., \cite{rv01} derived the hydrostatic equilibrium equation and the modified Chandrasekhar’s pulsation equation with in the framework of $f(R,T)$ gravity. In another work, Sarmah et al \cite{rv02} obtained the Chandrasekhar limiting masses as well as the dynamical instability criteria for white dwarfs in $f(R)$ gravitational theory. In this work, we discuss the stability using the sound speeds, i.e., (radial sound speed $\nu^{2}_{sr}$ and transverse sound speed $\nu^{2}_{st}$). These sound speeds can be defined as
\begin{equation}
\nu^{2}_{sr} =  \frac{dp_{r}}{dr} \times \frac{dr}{d\rho} =  \frac{dp_{r}}{d\rho},
\end{equation}
\begin{equation}
\nu^{2}_{st} =  \frac{dp_{t}}{dr} \times \frac{dr}{d\rho} =  \frac{dp_{t}}{d\rho}.
\end{equation}
According to Herrera's cracking condition \cite{ad17}, these velocities must lie between the interval [0,1]. The graphical analysis of radial sound speed is positive and decreasing in nature as seen in Fig. \ref{Fig.11}. Similarly, the transverse sound speed has similar nature as radial sound speed as shown in Fig. \ref{Fig.12}. It can also be noticed from these Figs. \ref{Fig.11} and \ref{Fig.12} that these sounds satisfy the Herrera condition, i.e., $0\leq\nu^{2}_{sr}$ and $\nu^{2}_{st}\leq1$. Abreu et al., \cite{ad18} discussed another condition to check either stellar structure is stable or unstable. The Abreu's condition is describe as $-1\leq \nu^{2}_{t} - \nu^{2}_{r} \leq 0$ and $0 \leq \nu^{2}_{r} - \nu^{2}_{t} \leq 1$. It can be observed from Fig. \ref{Fig.13} that the Abreu condition has also been satisfied.

\subsection{Energy Conditions}
\begin{figure}[h!]
\begin{tabular}{cccc}
\epsfig{file=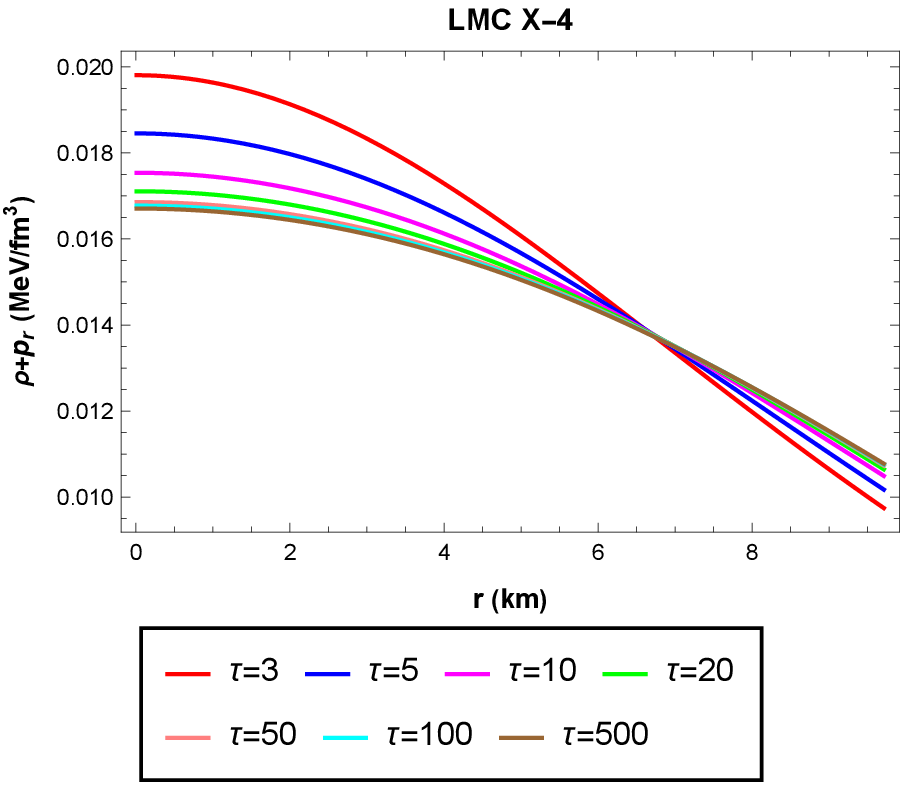,width=5.5cm,height=5.5cm}
\epsfig{file=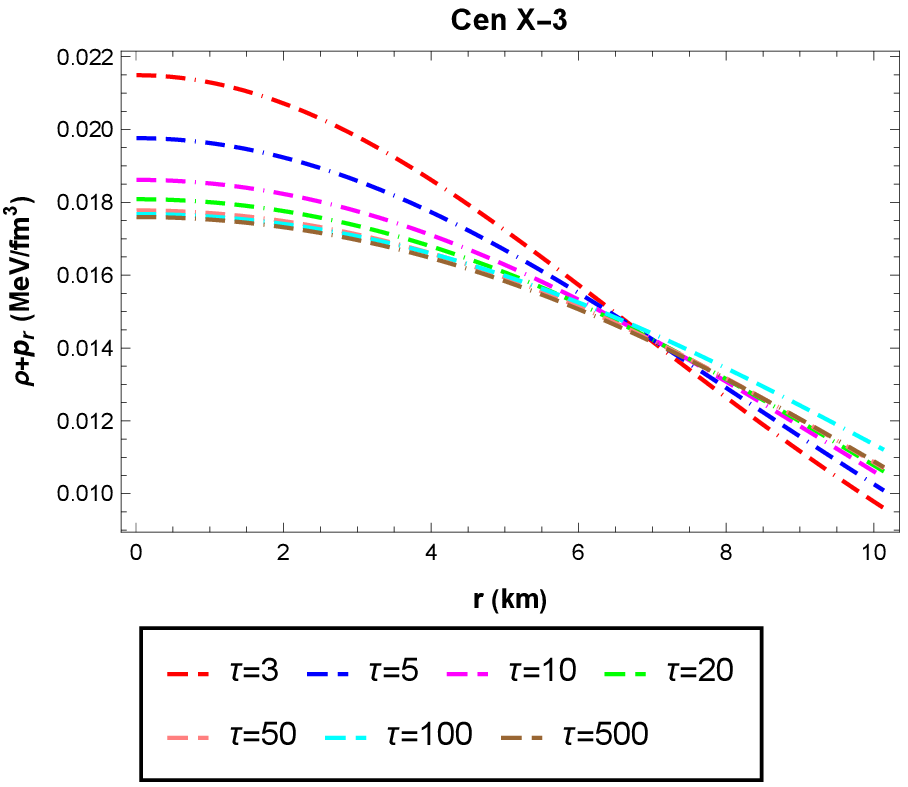,width=5.5cm,height=5.5cm}
\epsfig{file=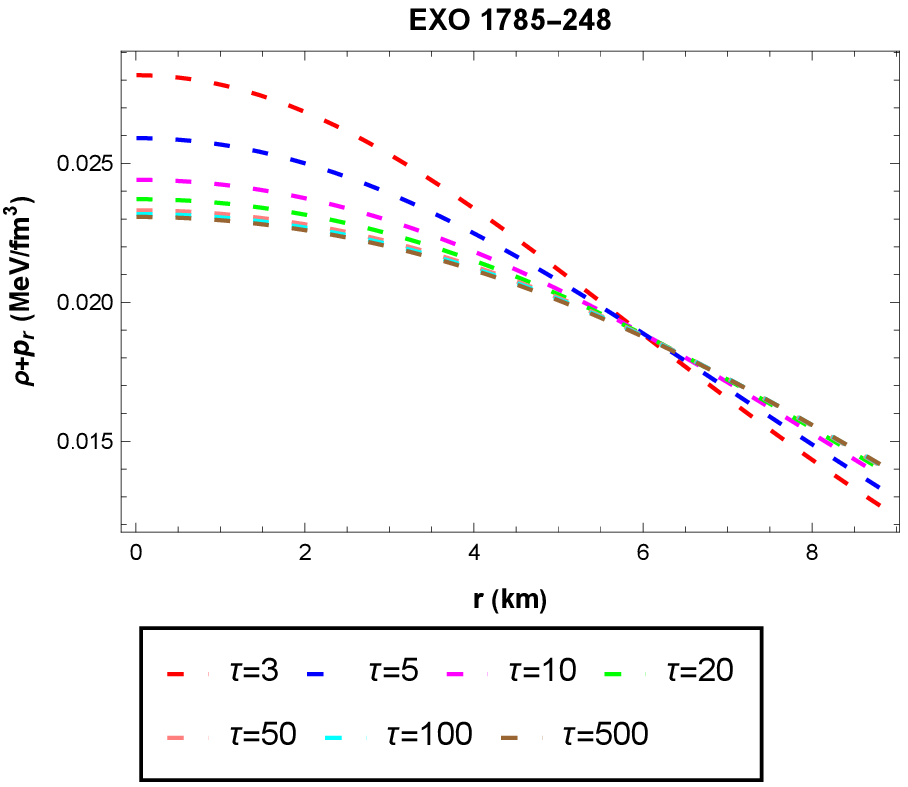,width=5.5cm,height=5.5cm}
\end{tabular}
\caption{\label{Fig.19} Graphical representation of $\rho+ p_{r}$ against radial coordinate.}
\end{figure}

Another essential condition for analyzing the stellar configuration is the energy conditions. These energy conditions are crucial in classifying the exotic and relativistic matter distributions in the stellar model. In the context of extended gravity theories, Atazadeh and Darabi \cite{ad23,ad23a,ad23b,ad23c} has been presented a formalism to check the validity of the energy conditions. These conditions can be described as
\begin{itemize}
  \item Null energy condition${(NEC) : \rho +p_{r} \geq 0, \rho +p_{t} \geq 0,}$
  \item Week energy condition$(WEC) : \rho \geq 0, \rho +p_{r} \geq 0, \rho +p_{t} \geq 0,$
  \item Strong energy condition$(SEC) : \rho +p_{r} \geq 0, \rho +p_{t} \geq 0, \rho +p_{r} +2p_{t} \geq 0,$
  \item Dominant energy condition${(DEC): \rho \geq |p_{r}|, \rho \geq |p_{t}|.}$
\end{itemize}
The graphical trend of $\rho+p_{r}$ is initially maximum at the core of the star and then becomes minimum as seen in Fig. \ref{Fig.19}. Similarly, $\rho+p_{t}$ has similar trends like $\rho+p_{r}$ as shown in Fig. \ref{Fig.19}. It can be concluded that NEC is satisfied due to positive nature of these two components i.e., $\rho+p_{r}$ and $\rho+p_{t}$. Furthermore, WEC is also satisfied because the graphical representation of energy density and NEC is satisfied. It can also be seen from Fig. \ref{Fig.22a} that the graphical analysis of $\rho+p_{r}+p_{t}$ is positive and decreasing in nature, which means that SEC is satisfied. It can also be noticed from Fig. \ref{Fig.21} and Fig. \ref{Fig.22} that DEC is also satisfied due to positive behavior of $\rho-p_{r}$ and $\rho-p_{t}$. It can be concluded that all the energy conditions are satisfied, which justifies that our star is viable.

\begin{figure}[h!]
\begin{tabular}{cccc}
\epsfig{file=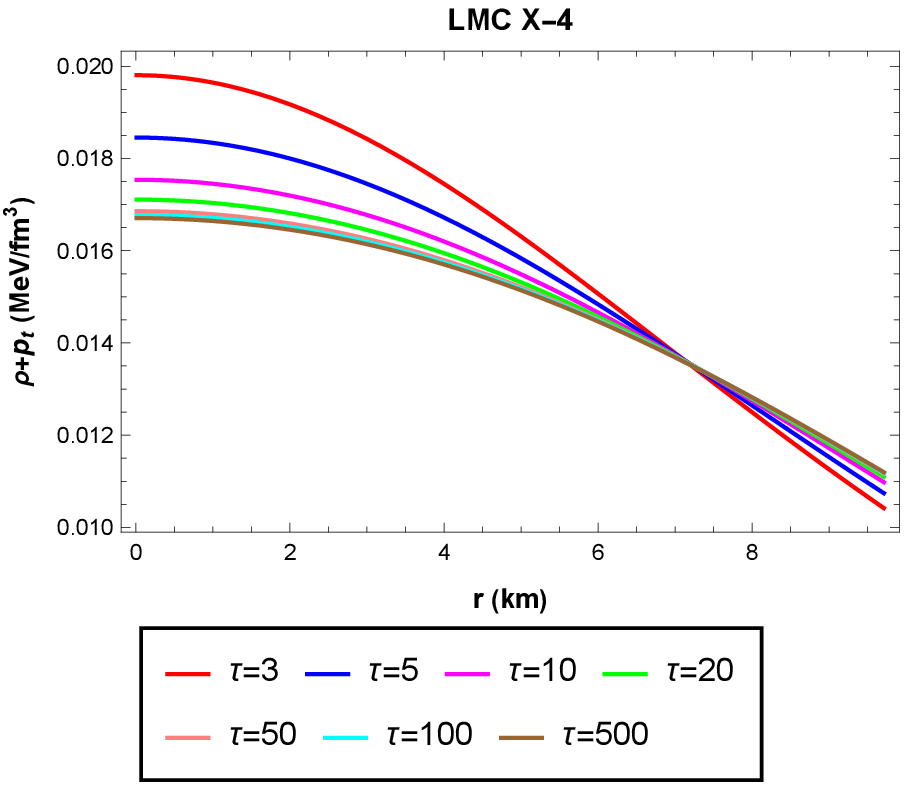,width=5.5cm,height=5.5cm}
\epsfig{file=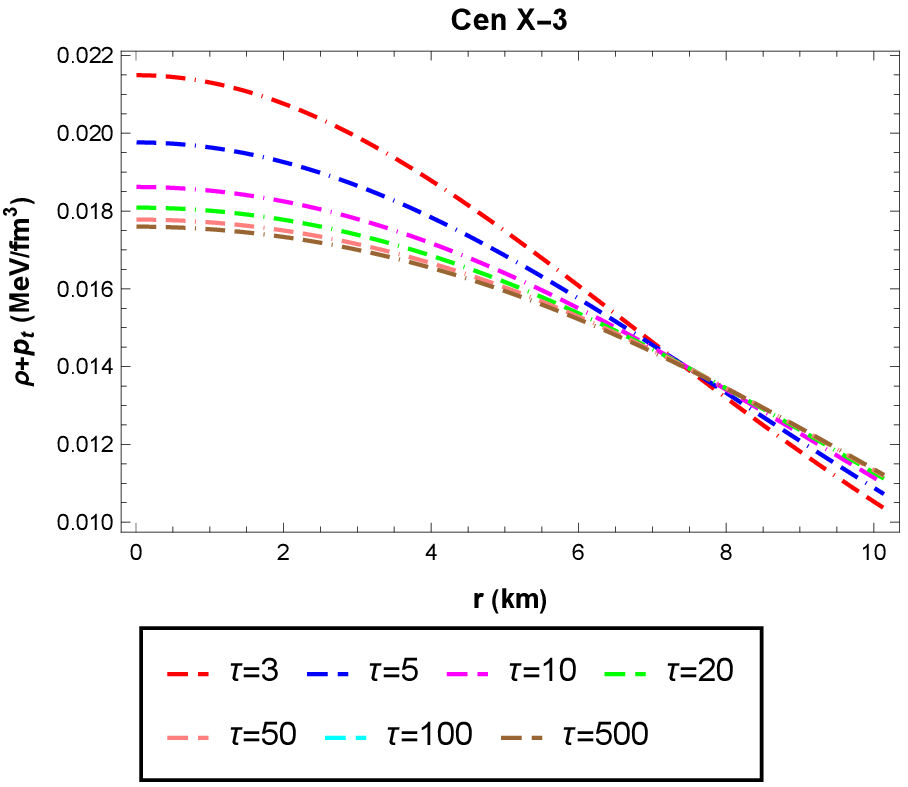,width=5.5cm,height=5.5cm}
\epsfig{file=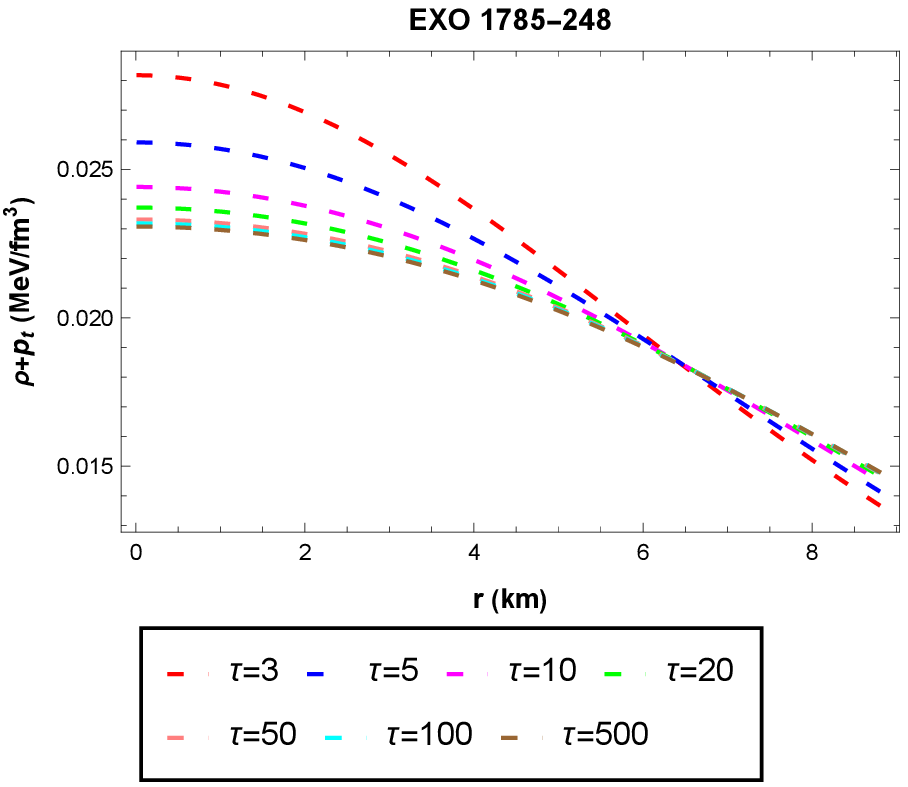,width=5.5cm,height=5.5cm}
\end{tabular}
\caption{\label{Fig.20} Graphical representation of $\rho+ p_{t}$ against radial coordinate.}
\end{figure}

\begin{figure}[h!]
\begin{tabular}{cccc}
\epsfig{file=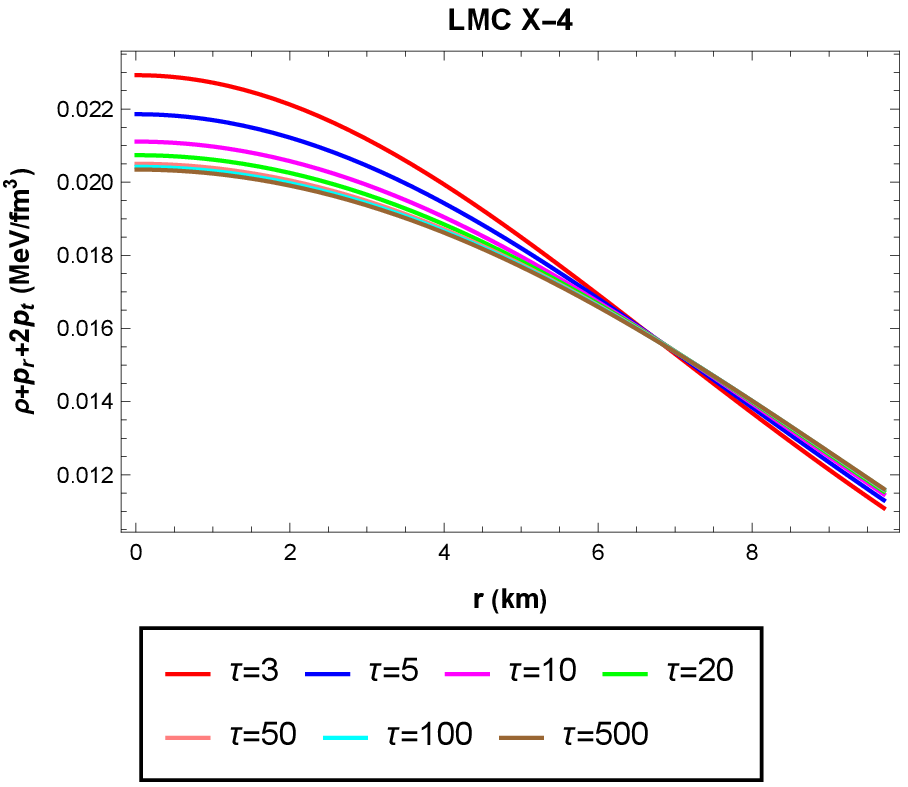,width=5.5cm,height=5.5cm}
\epsfig{file=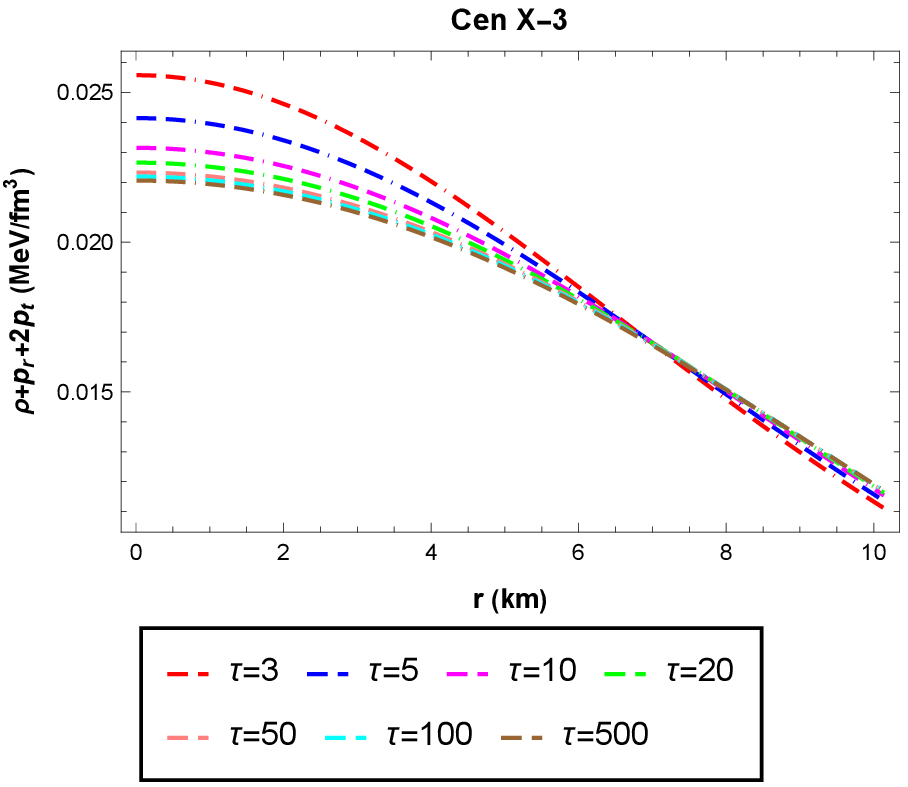,width=5.5cm,height=5.5cm}
\epsfig{file=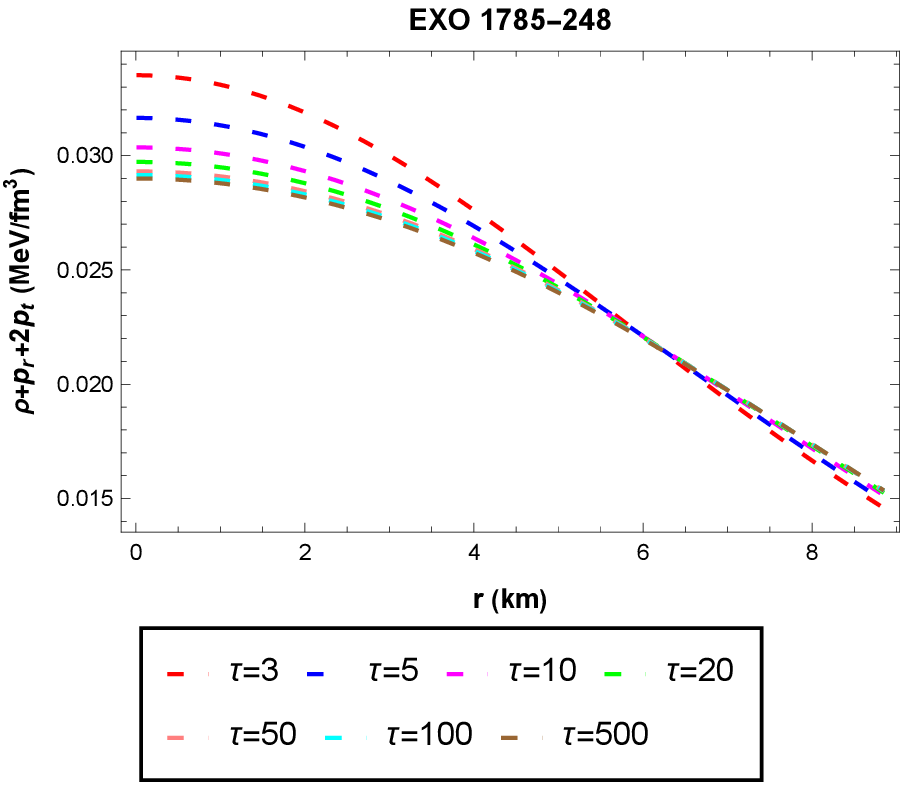,width=5.5cm,height=5.5cm}
\end{tabular}
\caption{\label{Fig.22a} Graphical representation of $\rho+ p_{r}+2p_{t}$ against radial coordinate.}
\end{figure}

\begin{figure}[h!]
\begin{tabular}{cccc}
\epsfig{file=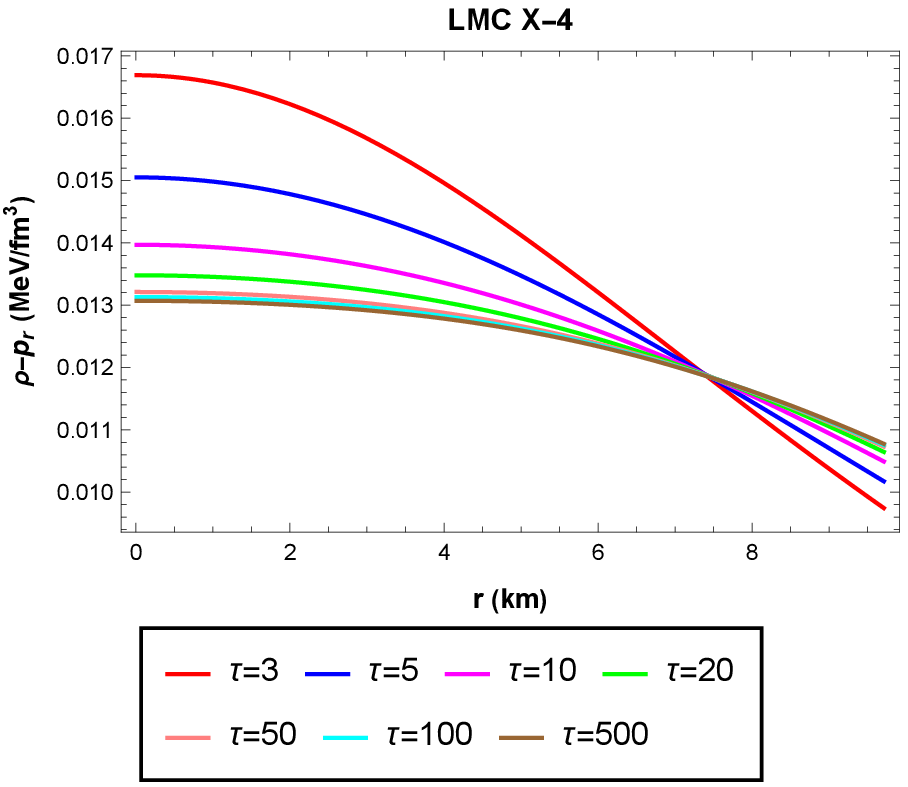,width=5.5cm,height=5.5cm}
\epsfig{file=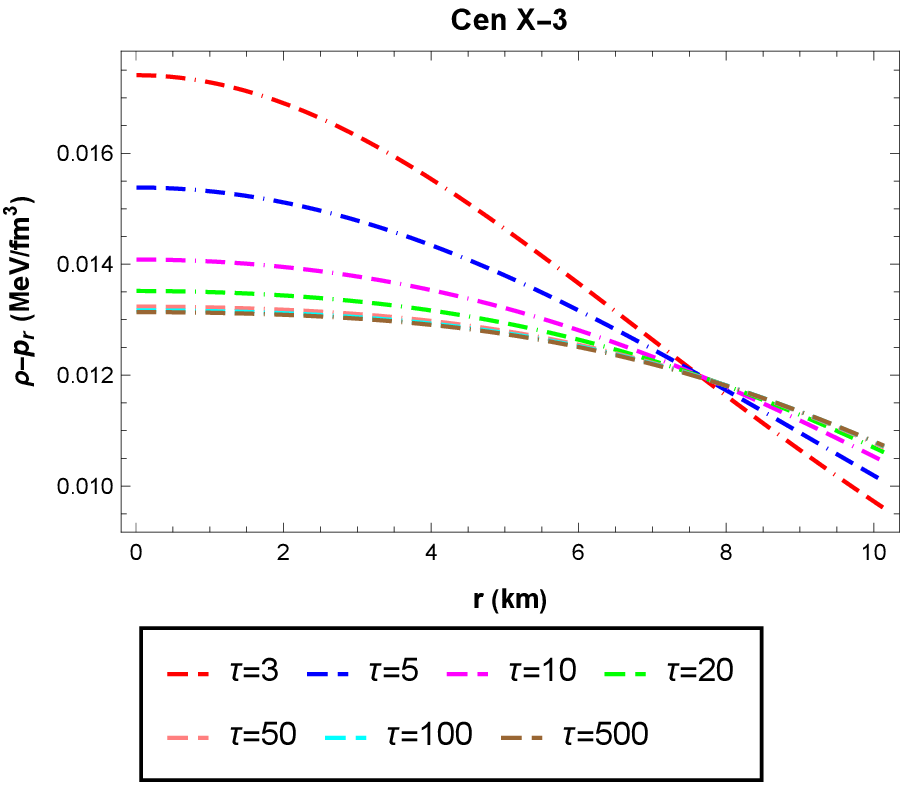,width=5.5cm,height=5.5cm}
\epsfig{file=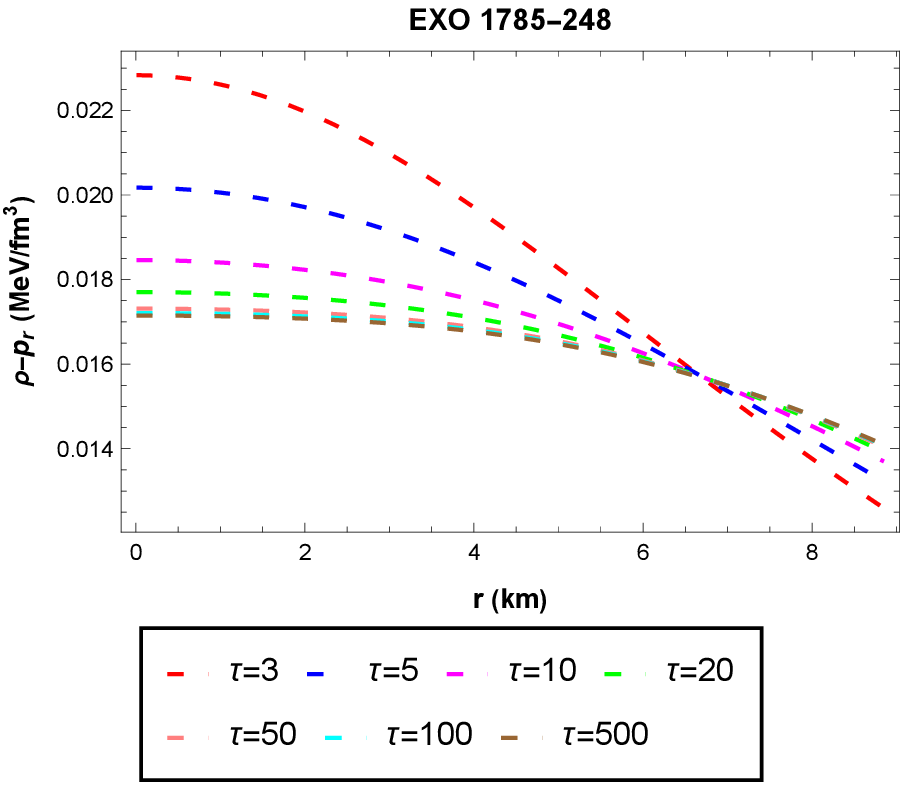,width=5.5cm,height=5.5cm}
\end{tabular}
\caption{\label{Fig.21} Graphical representation of $\rho- p_{r}$ against radial coordinate.}
\end{figure}

\begin{figure}[h!]
\begin{tabular}{cccc}
\epsfig{file=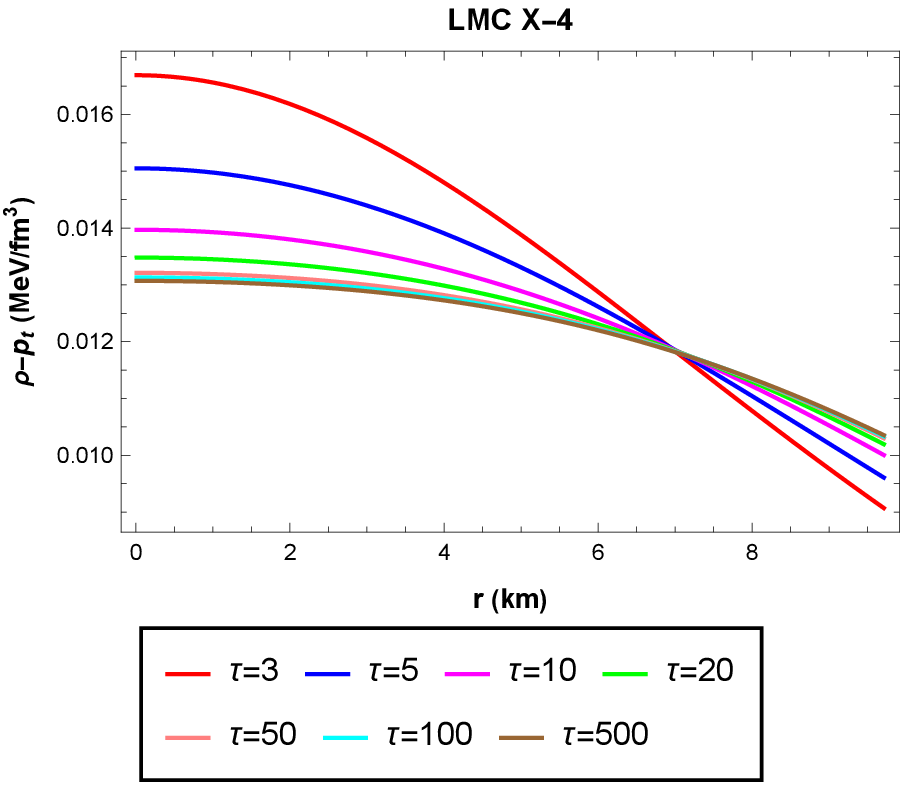,width=5.5cm,height=5.5cm}
\epsfig{file=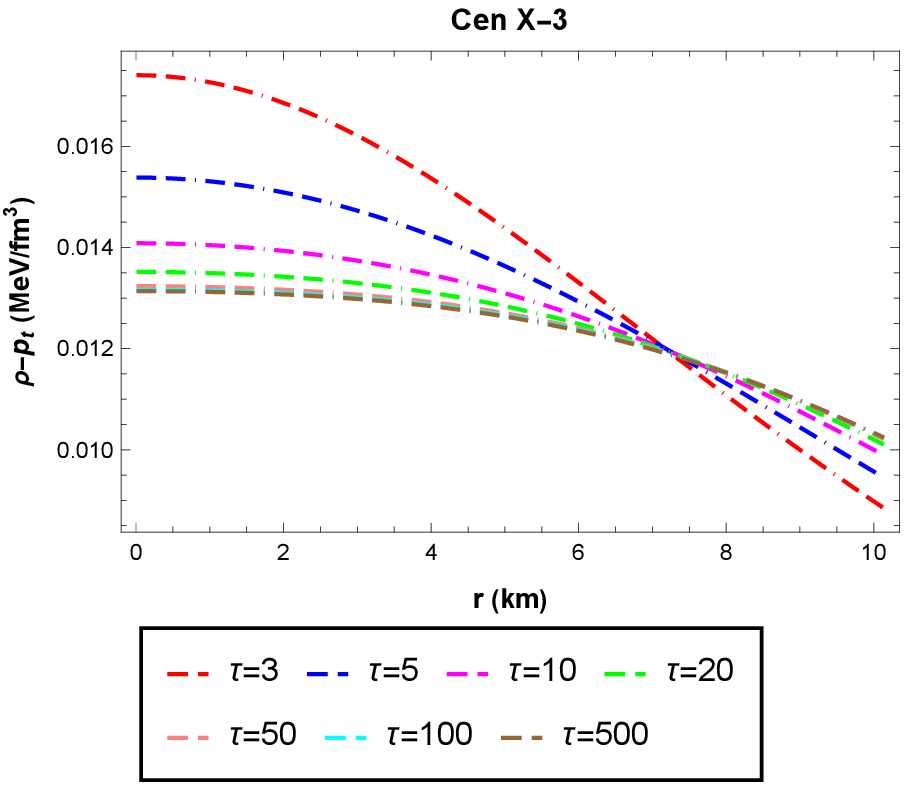,width=5.5cm,height=5.5cm}
\epsfig{file=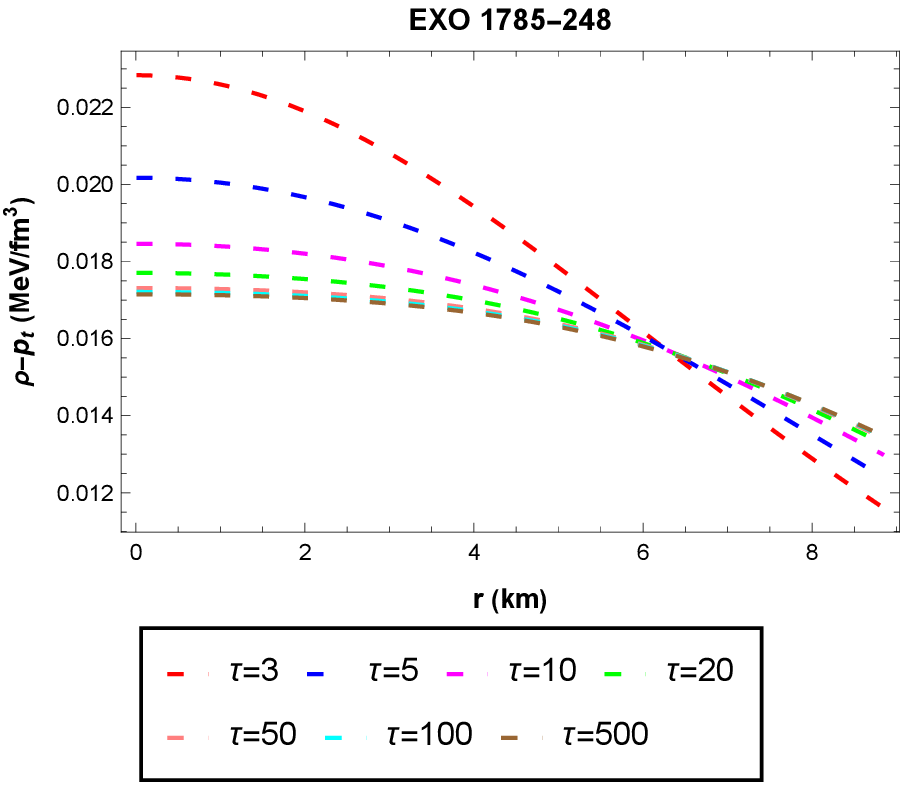,width=5.5cm,height=5.5cm}
\end{tabular}
\caption{\label{Fig.22} Graphical representation of $\rho- p_{t}$ against radial coordinate.}
\end{figure}

\subsection{Mass, Compactness factor and Surface Redshift}
\begin{figure}[h!]
\begin{tabular}{cccc}
\epsfig{file=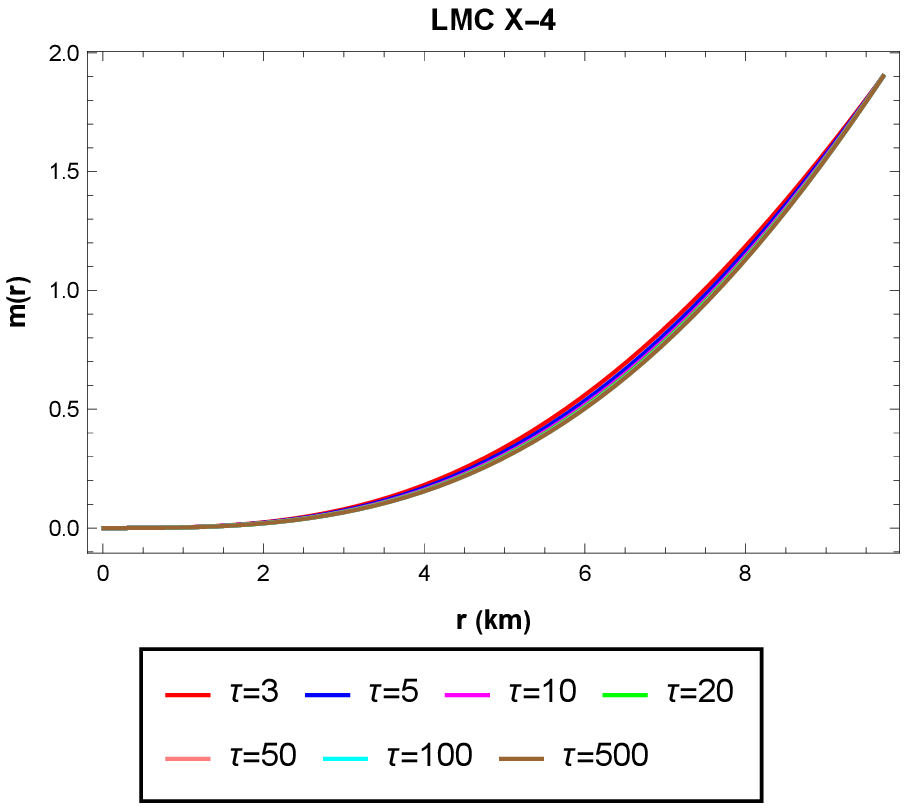,width=5.5cm,height=5.5cm}
\epsfig{file=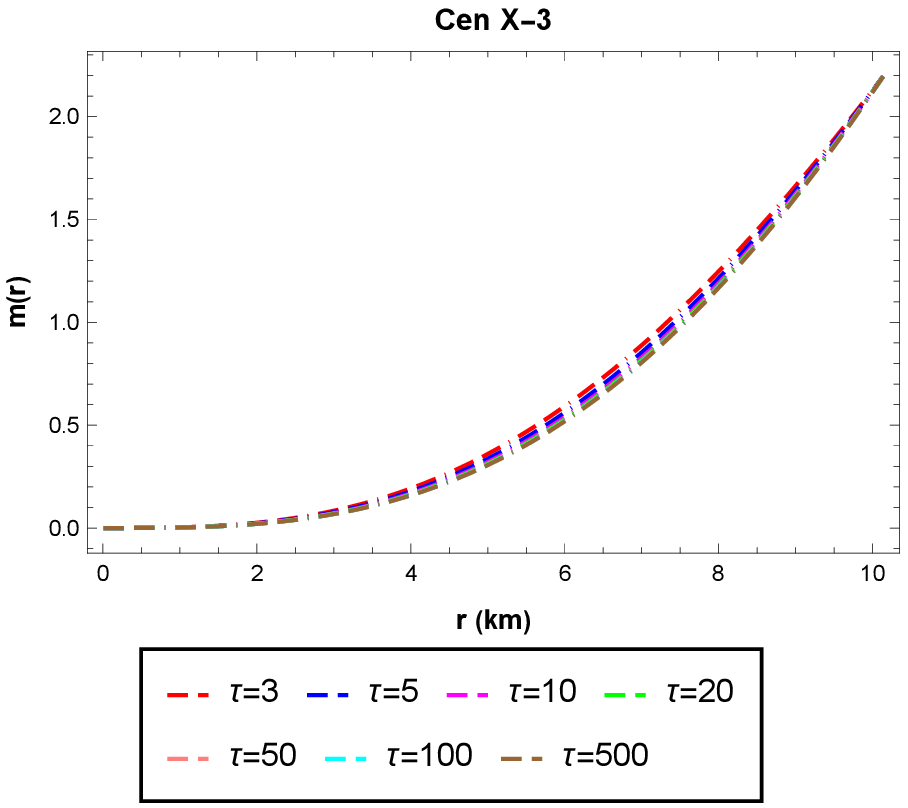,width=5.5cm,height=5.5cm}
\epsfig{file=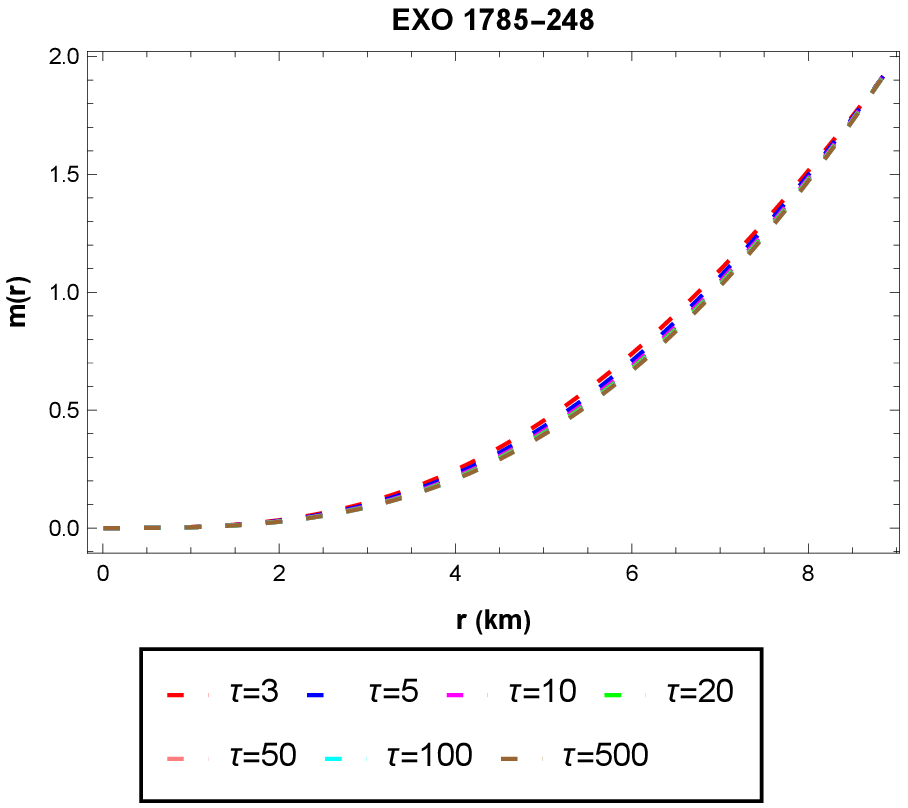,width=5.5cm,height=5.5cm}
\end{tabular}
\caption{\label{Fig.23} Graphical representation of mass functions against radial coordinate.}
\end{figure}

\begin{figure}[h!]
\begin{tabular}{cccc}
\epsfig{file=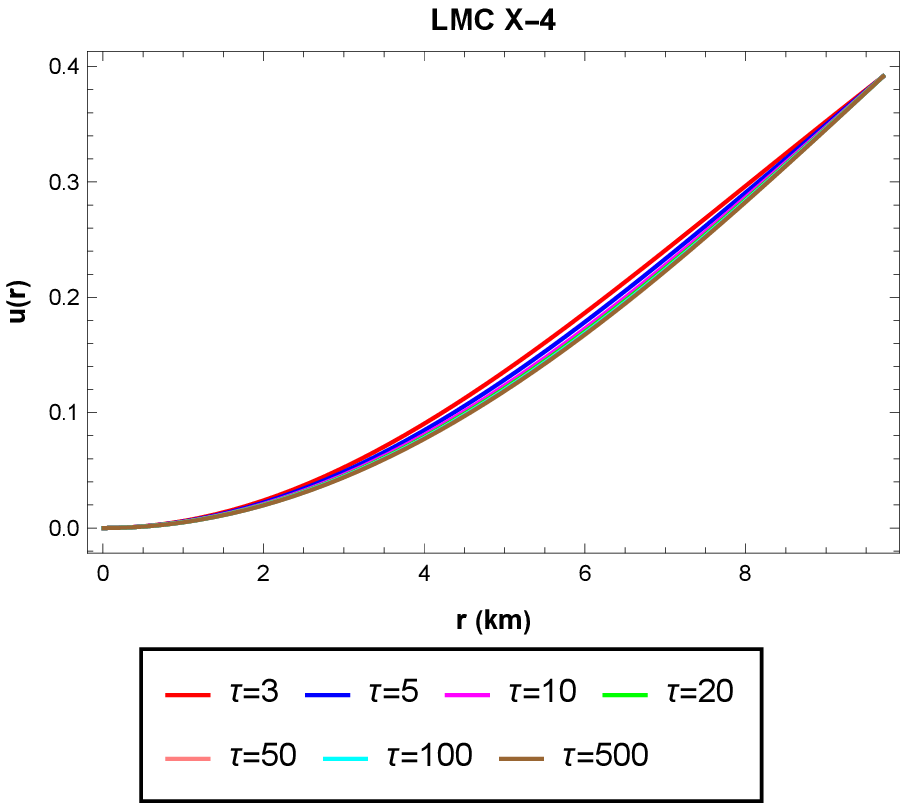,width=5.5cm,height=5.5cm}
\epsfig{file=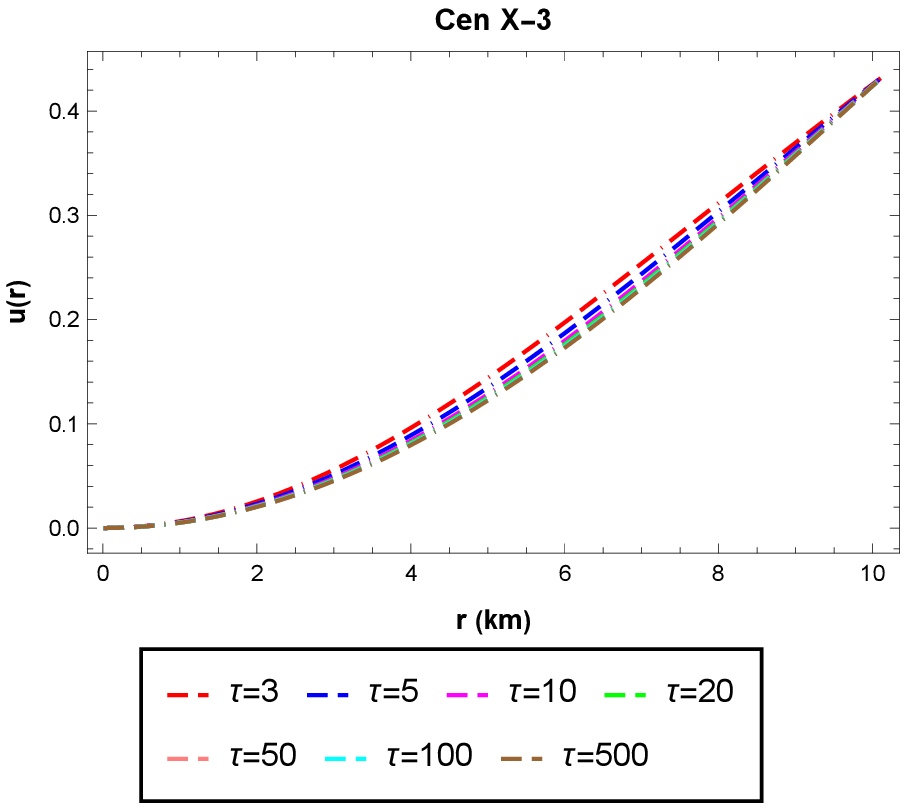,width=5.5cm,height=5.5cm}
\epsfig{file=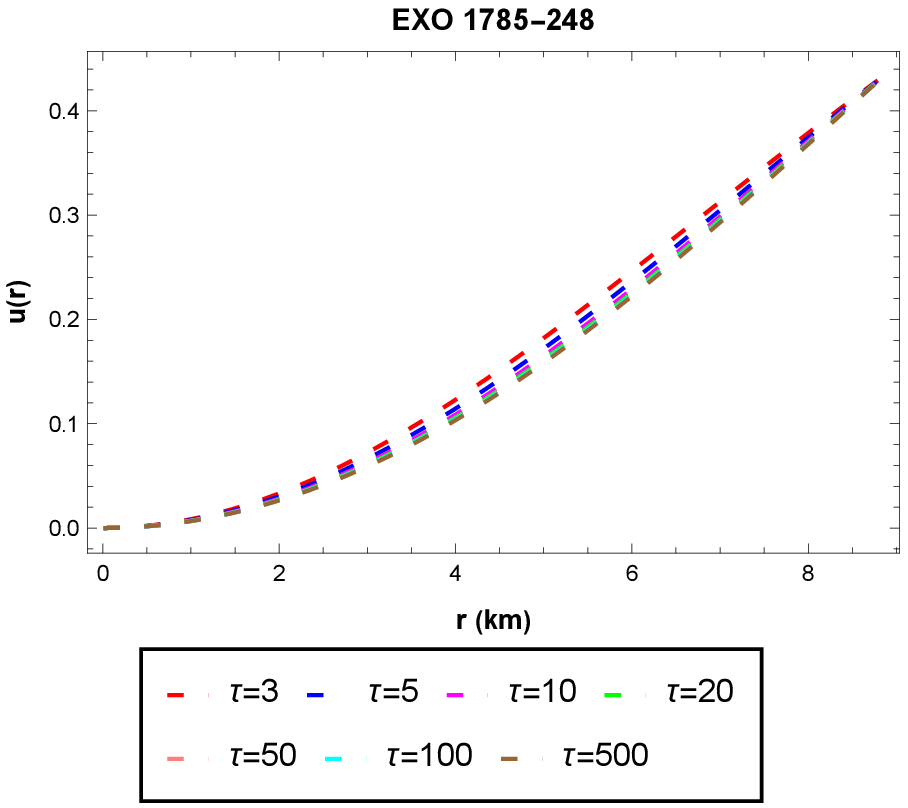,width=5.5cm,height=5.5cm}
\end{tabular}
\caption{\label{Fig.24} Graphical representation of compactness factor against radial coordinate.}
\end{figure}

\begin{figure}[h!]
\begin{tabular}{cccc}
\epsfig{file=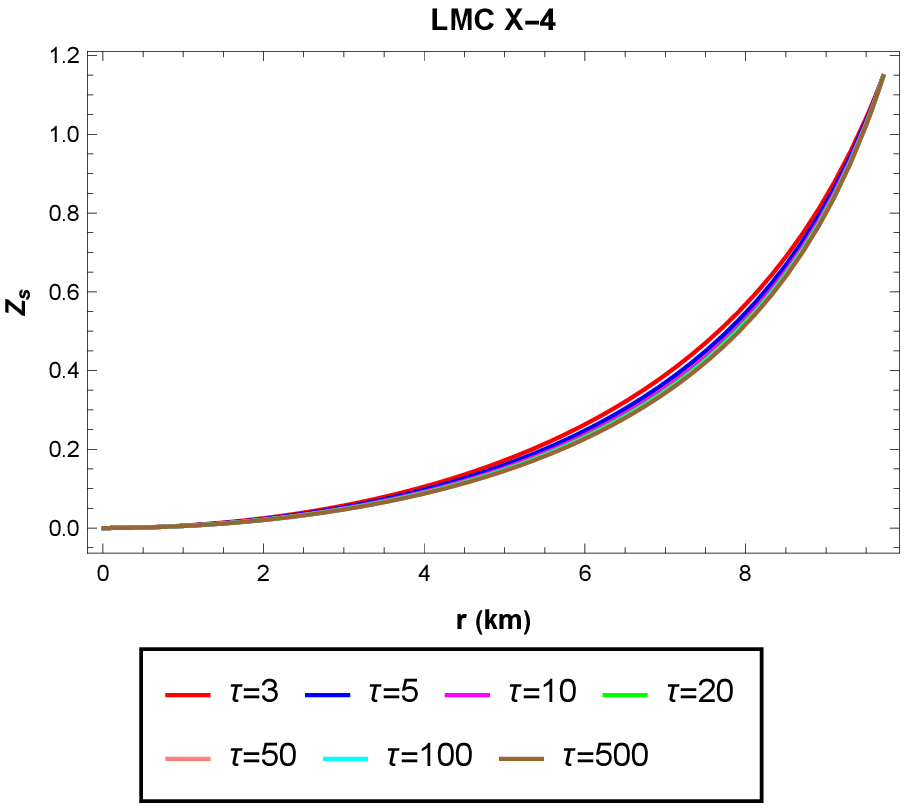,width=5.5cm,height=5.5cm}
\epsfig{file=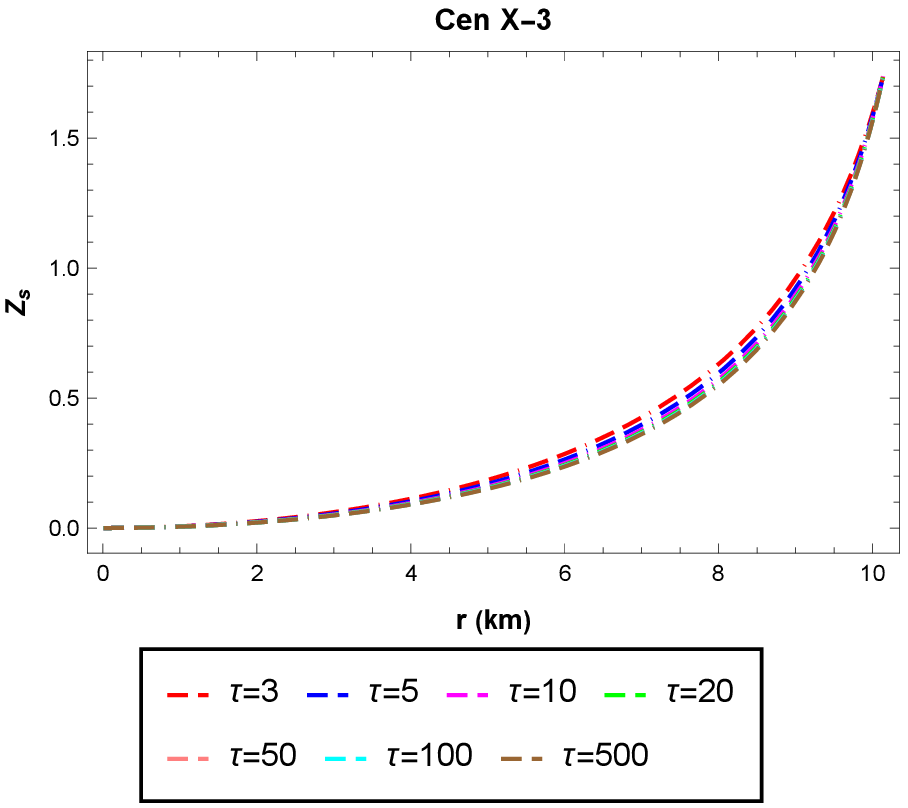,width=5.5cm,height=5.5cm}
\epsfig{file=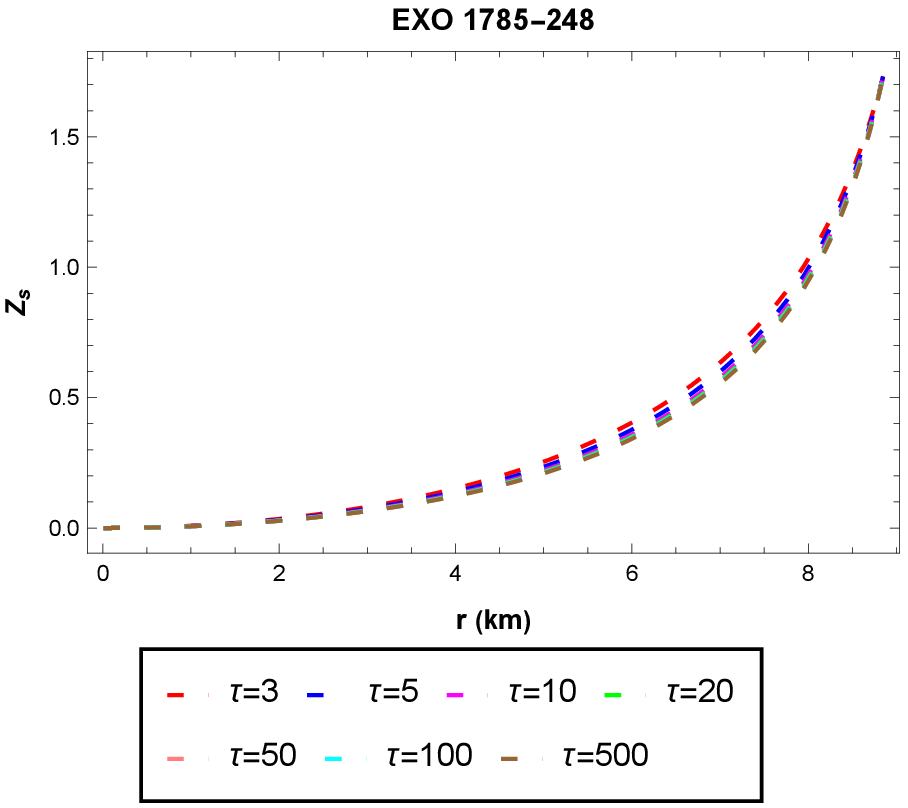,width=5.5cm,height=5.5cm}
\end{tabular}
\caption{\label{Fig.25} Graphical representation of surface redshift against radial coordinate.}
\end{figure}

The mass functions \cite{ad24} is defined as
\begin{equation}
 m(r) = 4\pi \int_{0}^{R}\rho r^{2}dr.
\end{equation}
It can be noticed that $M(r)\longrightarrow0$ as $r\longrightarrow0$, which represents that the mass function is regular at the center for the stellar object under the background of embedding approach in $f(R,\phi,X)$ theory of gravity as shown in Fig. \ref{Fig.23}. The compactness factor is denoted by $u(r)$ and defined as
\begin{equation}
 u(r) = \frac{m(r)}{r}.
\end{equation}
The graphical analysis of compactness factor can be seen in Fig. \ref{Fig.24}, which is positive and increasing in nature. One can also be noticed that compactness factor has similar nature like mass function shown in Fig. \ref{Fig.23}. It is believed that the Redshift function is also crucial component for the study of astrophysical objects and these physical features are
all linked with each other. The surface redshift function \cite{ad25} is given as
\begin{equation}
   Z_{s} +1 = [-2u(r) +1]^{\frac{-1}{2}}.
\end{equation}
The graphical plotting of surface redshift function is zero at the core and then becomes increasing when we move towards the boundary as sen in Fig. \ref{Fig.25}. Moreover, the graphical illustration redshift function is uniformly increasing, which confirms the stable behavior of the proposed $f(R,\phi,X)$ gravity models.

\section{Conclusion}
In this present study, we examined anisotropic stellar spheres in the $f(R,\phi, X)$ theory of gravity with embedding class-one approach. For the investigation of the stellar configuration, we use the compatible $f(R,\phi, X)=R +\gamma R^2+X-V(\phi)$ gravity model along with the spherically symmetric spacetime. These compact stars have different physical characteristics which are graphically examined for the present model by using different values of parameter $\tau$. Furthermore, we utilize the Schwarzschild's exterior geometry to find the unknown constants. All the essential results that are satisfied in a stellar configuration are summarized as follows:
\begin{itemize}
\item The graphical analysis of energy density and tangential pressure is positive, decreasing and maximum at the core throughout the internal configuration. Similarly, the graphical analysis of radial pressure is maximum at the center of stellar structure and ultimately vanishes at the boundary of the star. The behavior of energy density and pressure components suggest the high compactness of the center of the star, which represent that our model under inspection is feasible for the exterior region of the core.

\item The derivative of energy density, radial pressure and tangential pressure with respect to radial component show negative, which demonstrates that all these functions are positive and attained maximum output at the center of the star, which is again a valid condition for the compact star.

\item There is an interesting fact that if $\Delta=0$, then there is isotropic pressure in the matter distribution. If anisotropic measurement is positive $\Delta>0$, then anisotropic force is outward. On the other hand, force is inward if an anisotropic measurement is negative ($\Delta<0$). In our current manuscript, the anisotropy is positive, directed outward and zero at the core of the considered stars as seen in Fig. \ref{Fig.7}.

\item The graphical behavior of radial and tangential EoS parameters is maximum at core of star, monotonically decreasing towards the boundary, and lying between 0 and 1, which is again a crucial condition of stellar structure.

\item From Fig. \ref{Fig.10}, it can be observed that the behavior of metric potentials is finite, positive at the center, and increasing towards the boundary, which shows that the present stellar structures are free from singularity.

\item The graphical analysis of radial sound speed and transverse sound speed is positive and decreasing in nature as seen in Fig. \ref{Fig.11} and Fig. \ref{Fig.12} and satisfy the Herrera condition, i.e., $0\leq\nu^{2}_{sr}$ and $\nu^{2}_{st}\leq1$. It can be noticed that the Abreu condition i.e., $-1\leq \nu^{2}_{t} - \nu^{2}_{r} \leq 0$ and $0 \leq \nu^{2}_{r} - \nu^{2}_{t} \leq 1$ is also satisfied for considered stars.

\item All the energy conditions namely NEC, WEC, SEC and DEC are satisfied, which justifies that our star is viable.

\item It can be noticed that $M(r)\longrightarrow0$ as $r\longrightarrow0$, which represents that the mass function is regular at the center for the stellar object under the background of embedding approach in $f(R,\phi,X)$ theory of gravity as shown in Fig. \ref{Fig.23}.

\item The graphical analysis of compactness factor can be seen in Fig. \ref{Fig.24}, which is positive and increasing in nature. One can also be noticed that compactness factor has similar nature like mass function shown in Fig. \ref{Fig.23}.

\item The graphical plotting of surface redshift function is zero at the core and then becomes increasing when we move towards the boundary as sen in Fig. \ref{Fig.25}. Moreover, the graphical illustration redshift function is uniformly increasing, which confirms the stable behavior of the proposed $f(R,\phi,X)$ gravity models.
\end{itemize}
Modified $f(R,\phi,X)$ theory of gravity plays an alluring role in the study of stellar structures. Hence, we conclude that our $f(R,\phi,X)$ model in the frame of anisotropic fluid is stable and consistent, as all physical characteristics of compact objects obey physically obtained patterns. Moreover, it is mentioned here that our results are more identical to the past relevant work in $f(R)$ gravity \cite{a16}.

\section*{Data Availability Statement}
\hskip\parindent
\small
No data was used for the research in this article. It is pure mathematics.

\section*{Conflict of Interest}
\hskip\parindent
\small
The authors declare that they have no conflict of interest.

\section*{Contributions}
\hskip\parindent
\small
 We declare that all the authors have same contributions to this paper.

\section*{Data Availability Statement}
The authors declare that the data supporting the findings of this study are available within the article.

\section*{Acknowledgement}
Adnan Malik acknowledges the Grant No. YS304023912 to support his Postdoctoral Fellowship at Zhejiang Normal University, China.
 This paper was completed during the postdoctoral fellowship of the first author under the supervision of Professor Xia at Zhejiang Normal University, China.

\section*{References}

\section{Appendix}
\begin{eqnarray}
t_1(r)&=&(1+hr^2).\nonumber\\
t_2(r)&=&(t_1(r)^2+ohr^2t_1(r)^{\tau}).\nonumber\\
t_3(r)&=&60-3\tau+2(-20+\tau(19+3\tau))hr^2+(1+\tau)(-4+\tau+\tau^2)h^2r^4.\nonumber\\
t_4(r)&=&5+hr^2(55-20\tau+(47+4(-4+\tau)\tau)hr^2+(29-12\tau(1+\tau))h^2r^4).\nonumber\\
t_5(r)&=&3(1+\tau)r^2+(120+\tau(-312+241\tau))\gamma-3r^{2+2\alpha}\alpha^2.\nonumber\\
t_6(r)&=&(4+\tau)r^2+12(-10+9\tau)\gamma-2r^{2+2\alpha}\alpha^2.\nonumber\\
t_7(r)&=&3\tau r^2+216\gamma+2\tau(-70+\tau(-87+71\tau))\gamma-2r^{2+2\alpha}\alpha^2.\nonumber\\
t_8(r)&=&(-1+2\tau)r^2+2(-24+\tau(56+\tau(-27+\tau(-30+17\tau))))\gamma-r^{2+2\alpha}\alpha^2.\nonumber\\
t_9(r)&=&(20-3r^{2\alpha}\alpha^2+2h^2r^216(4+3\tau)r^2+420\gamma-16\tau(3+14\tau)\alpha-9r^{2+2\alpha}\alpha^2)+4h((14+3\tau)r^2+4(27-46\tau)\alpha-\nonumber\\
&=&3r^{2+2\alpha}\alpha^2)+4h^3r^4((2+9\tau)r^2+4(27+\tau(4(5-16\tau)\tau))\gamma-3r^{2+2\alpha}\alpha^2)+h^4r^6(4(-1+3\tau)r^2+4(75+4\tau(-36\nonumber\\
&=&+\tau(21+2\tau-6\tau^2)))\gamma-3r^{2+2\alpha}\alpha^2).\nonumber\\
t_{10}(r)&=&(1+(-1+\tau)hr^2).\nonumber\\
t_{11}(r)&=&116+4(-3+56\tau)hr^2+(-84+\tau(16+155\tau))h^2r^4+(-1+\tau)(-44+\tau(68+39\tau))h^3r^6.\nonumber\\
t_{12}(r)&=&33-28\tau+(-15+4(20-3\tau)\tau)hr^2+(-5+4\tau(-1+9\tau))h^2r^4.\nonumber\\
t_{13}(r)&=&8+66\tau+(-12+\tau(16+59\tau))hr^2+(24+7\tau(6+\tau(-1+4\tau)))h^2r^4+2(-1+\tau)(1+\tau(11+\tau(4+3\tau)))h^3r^6.\nonumber\\
t_{14}(r)&=&(4+4(1+2\tau)hr^2+16(-1+4\tau)h\gamma-r^{2\alpha}t_{1}(r)\alpha^2).\nonumber\\
t_{15}(r)&=&r^2t_1(r)^6+o^3h^3r^8t_1(r)^{3\tau}+2t_1(r)^4(11+3hr^2(-18+hr^2))\gamma+o^2h^2r^4t_1(r)^{2\tau}(3(r+hr^3)^2+4(59+hr^2(4\nonumber\\
&=&-39hr^2))\gamma)+ohr^2t_1(r)^{2+\tau}(3(r+hr^3)^2+h(-231+hr^2(194+45hr^2))\gamma).\nonumber\\
t_{16}(r)&=&(t_1(r)^7+8o^4h^4r^6(t_1(r)^4)^{\tau}\gamma+8ht_1(r)^5(-5+hr^2(8+hr^2))\gamma-oh(1+hr^2)^{3+\tau}(-3(r+hr^3)^2+4(\tau+hr^2(-1\nonumber\\
&=&+hr^2(-52+17hr^2)))\gamma)+o^3h^3r^4t_1(r)^{3\tau}(26\gamma+r^2t_1(r)+6h(6+7hr^2)\gamma))+o^3h^2r^2t_1(r)^{2+2\tau}(226\gamma+r^2(3-\nonumber\\
&=&80h\alpha+hr^2(3+158h\gamma))).\nonumber\\
t_{17}(r)&=&-102\gamma+r^2(-3+3hr^2(-1+hr^2+h^2r^4)-2h(129+hr^2(89+43hr^2))\gamma+r^{2\alpha}t_1(r)^3\alpha^2).\nonumber
\end{eqnarray}

\end{document}